\pdfoutput=1
\documentclass[pra,aps,showpacs,groupedaddress,superscriptaddress,twocolumn,10pt]{revtex4-1}
\usepackage{amssymb}
\usepackage[english]{babel}
\usepackage{graphicx}
\usepackage{dcolumn}
\usepackage{bm}
\usepackage{color}

\begin{document}

\title{Optical lattices as a tool to study defect-induced superfluidity}

\author{Grigory E. Astrakharchik}
\affiliation{Departament de F\'{\i}sica, Universitat Polit\`{e}cnica de Catalunya, 08034 Barcelona, Spain}

\author{Konstantin V. Krutitsky}
\affiliation{Fakult\"at f\"ur Physik der Universit\"at Duisburg-Essen, Campus Duisburg, 47048 Duisburg, Germany }

\author{Maciej Lewenstein}
\affiliation{ICFO -- Institut de Ci\`{e}ncies Fot\`{o}niques, The Barcelona Institute of Science and Technology, 08860 Castelldefels, Spain}
\affiliation{ICREA, Lluis Companys 23, 08010 Barcelona, Spain}

\author{Ferran Mazzanti}
\affiliation{Departament de F\'{\i}sica, Universitat Polit\`{e}cnica de Catalunya, 08034 Barcelona, Spain}

\author{Jordi Boronat}
\affiliation{Departament de F\'{\i}sica, Universitat Polit\`{e}cnica de Catalunya, 08034 Barcelona, Spain}

\date{v1 December 22, 2016; v2 September 6, 2017}


\begin{abstract}
We study the superfluid response, the energetic and structural properties of a one-dimensional ultracold Bose gas in an optical lattice of arbitrary strength.
We use the Bose-Fermi mapping in the limit of infinitely large repulsive interaction and the diffusion Monte Carlo method in the case of finite interaction.
For slightly incommensurate fillings we find a superfluid behavior which is discussed in terms of vacancies and interstitials. It is shown that both the excitation spectrum and static structure factor are different for the cases of microscopic and macroscopic fractions of defects.
This system provides a extremely well-controlled model for studying defect-induced superfluidity.
\end{abstract}

\maketitle

\section{Introduction}

A supersolid is a spatially ordered material with superfluid properties.
Practically, since the discovery of superfluidity by P.~Kapitza, J.~F.~Allen and D.~Misener~\cite{bookSF,KhalatnikovBook,PitaevskiiStringariBook}, there have been constant efforts to understand and find systems that exhibit supersolidity.
From the theoretical point of view, the most plausible mechanism of supersolidity is based on the so called Andreev-Lifshitz-Chester scenario~\cite{AndreevL,Chester70}.
In this theory it is assumed that vacancies, empty sites normally occupied by particles in an ideal crystal, exist even at absolute zero.
These vacancies might be caused by quantum fluctuations, which also causes them to move from site to site.
Because vacancies are bosons, if such clouds of vacancies can exist at very low temperature $T$, then a Bose-Einstein condensation of vacancies could occur at temperatures less than a few tenths of a kelvin.
A coherent flow of vacancies is equivalent to a ``superflow'' (frictionless flow) of particles in the opposite direction.
Despite the presence of the gas of vacancies, the ordered structure of a crystal is maintained, although with less than one particle on each lattice site on average~\cite{Chan2008,Boninsegni2012,Rota12}.
Shevchenko presented another scenario for supersolidity, in which mass flow occurs along dislocation cores forming a three-dimensional (3D) networks~\cite{Shevchenko1987,Boninsegni2007,Pollet2008}.
This model has many difficulties, since it requires large density of dislocations and their interconnections.
Supersolidity regarded from this point of view is inevitably connected to the question whether solid helium is a quantum supersolid (for a recent review see Ref.~\cite{Balibar2016}).

Leggett~\cite{Leggett1970} was perhaps the first to propose various direct tests of supersolidity, but so far there is no clear evidence for supersolidity in the ground state ($T=0$) of superfluid $^4$He, and there is an ongoing
debate~\cite{Kim2004,Kim2004b,Todoshchenko2007,Day2007,Reppy2010,Maris2012,KimChan2012,Hallock2015}.
The debate concerns the effects of elasticity, the presence of defects due to preparation, or the absence of vacancies at $T=0$ (cf.~\cite{Sanders2007,Weiss2004,Witze2010}).
P.W. Anderson argued~\cite{Anderson2007,Anderson2008,Anderson2009} that Bose fluids above the critical temperature may behave as incompressible vortex fluids and this may explain recent experimental results concerning ``supersolidity''.

Recent intensive studies of helium crystals by Balibar's group, on the other hand, suggest the following:
(i) no supersolidity in bulk $^4$He crystals;
(ii) no clear evidence for superfluidity in defects of $^4$He crystals except from simulations;
(iii) Clear evidence for quantum tunneling of kinks on dislocations, and $^3$He impurities in bulk $^4$He.

Defects in a quantum crystal might lead to a finite superfluid signal, while the system can not be considered as being in a true ground state as a crystal lattice without defects would have a smaller energy.
Quite similarly, the Andreev-Lifshitz-Chester mechanism is driven by the presence of defects, although in the true ground state.
We find it appealing to study how a finite concentration of defects induces the superfluid signal in the system.
We believe that optical lattices provide an ideal setup for such studies, since they permit us to:
\begin{itemize}
\item create a ground state with vacancies (not possible in helium);

\item create a microscopic number of defects, by removing some atoms using atom-microscope techniques~\cite{Greiner2009,Kuhr2010,Kuhr2015}
(which should also be a feasible scenario for helium);
\item create a macroscopic number of defects by appropriate loading (not feasible regime for helium).
\end{itemize}
In the last case, a mismatch between two characteristic momenta -- the Fermi momentum and that corresponding to the edge of the first Brillouin zone -- is expected to result in intricate spatial correlations.

In this paper we investigate this possibility and study a one-dimensional Bose gas with short-range repulsive interactions
in an optical lattice at incommensurate filling.
We confirm that it provides a perfect candidate to study the defect-induced superfluidity, similarly to the Andreev-Lifshitz-Chester scenario for supersolidity.
We demonstrate how the presence of vacancies leads to the appearance of superfluidity with modulated density.
Of course, modulation in the present study is not spontaneous -- it is caused by the presence of the optical lattice.
We expect, however, that the same mechanism will be effective in lattice gases with dipolar or soft core potentials, and will be responsible for formation of supersolids at low commensurate fillings, leading to a spontaneous appearance of density modulations at periodicities larger than the lattice constant (cf.~\cite{Goral2002,Barbara2010}).
The detailed analysis of this mechanism in dipolar gases will be discussed elsewhere.

The paper is organized as follows.
In Section~\ref{Sec:H} we introduce the model Hamiltonian.
We begin by discussing the case of infinite interactions (Tonks-Girardeau limit) in Section~\ref{Sec:TG} and analyze the single-particle excitation spectrum in a homogeneous system in Section~\ref{Sec:TG:hom} and in the presence of an optical lattice in Section~\ref{Sec:TG:lattice}.
We proceed then to the case of finite interactions in Section~\ref{Sec:finite interaction}.
We stress here the fundamental differences between excitations in the absence and in the presence of defects (vacancies and interstitials).
Section~\ref{Sec:sf} is devoted to the discussion of the superfluid fraction, using the winding number technique, also known as boost method (cf.~\cite{Lieb2002,Roth2003,Damski2003}).
In Section~\ref{Sec:Sk} we turn to the discussion of the static structure factor which is useful for understanding both structural properties and excitations.
The particle-particle and vacancy-vacancy pair distribution function are presented in Sections~\ref{Sec:g2} and~\ref{Sec:g2vac}.
Finally, we summarize our results and draw the main conclusions in Section~\ref{Sec:conclusions}.

\section{Hamiltonian\label{Sec:H}}

We consider $N$ ultracold atoms in a one-dimensional optical lattice created by a largely detuned laser field.
This gives rise to an external periodic potential
\begin{eqnarray}
\label{VL}
V_{\rm L}(x)
=
V_0
\cos^2
\left(
     \pi \frac{x}{a_0}
\right)
\end{eqnarray}
of strength $V_0$ with lattice constant $a_0$.
A characteristic energy associated with the lattice is the {\it recoil energy} $E_{\rm rec} = \pi^2\hbar^2/(2ma_0^2)$, which we will use as a unit of energy.

The atoms are assumed to be bosons  of mass $m$ interacting with each other by a contact potential of strength $g_{\rm 1D}=-2\hbar^2/(m a_{\rm 1D})$, with $a_{\rm 1D}<0$ being the one-dimensional $s$-wave scattering length.
The first quantization form of the Hamiltonian of the system then reads
\begin{equation}
\label{Eq:H:continuous}
\hat H
=
\sum_{i=1}^N
\left[
    -{\hbar^2 \over 2m}
    {\partial^2 \over \partial x_i^2}
    +
    V_{\rm L}(x_i)
\right]
+
g_{\rm 1D}
\sum_{i<j}
\delta(x_i-x_j)
\;,
\end{equation}
where $x_i$, with $i=1,\dots,N$ are the particle coordinates.
We impose periodic boundary conditions on a box of size $La_0$, where $L$ is an integer.

The ground-state properties of Hamiltonian~(\ref{Eq:H:continuous}) are studied using the diffusion Monte Carlo (DMC)  algorithm~\cite{Casu95} which solves the Schr\"odinger equation in imaginary time.
The DMC method gives an exact estimation (in statistical sense) of any observable commuting with the Hamiltonian, and delivers bias-free predictions for other observables by means of pure estimator techniques~\cite{Casu95}.

In deep optical lattices ($V_0/E_{\rm rec}\gg1$), the Hamiltonian~(\ref{Eq:H:continuous}) reduces to the Bose-Hubbard model.
In its standard and simplest form, the Hamiltonian of this model, in second quantization, is given by
\begin{equation}
\hat H_{\rm BH}
=
-J
\sum_{\ell=1}^L
\left(
    \hat a_\ell^\dagger
    \hat a_{\ell+1}^{\phantom{\dagger}}
    +
    {\rm h.c.}
\right)
+
\frac{U}{2}
\sum_{\ell=1}^L
\hat a_\ell^\dagger
\hat a_\ell^\dagger
\hat a_\ell^{\phantom{\dagger}}
\hat a_\ell^{\phantom{\dagger}}\;,
\label{Eq:H:BH}
\end{equation}
with $\hat a_\ell^\dagger$, $\hat a_\ell$ standard particle creation and annihilation operators.
The tunneling matrix element $J$ and the interaction constant $U$ are determined as~\cite{Lewenstein12}
\begin{eqnarray}
\label{Eq:JU}
J
&=&
-\int_{0}^{La_0}
W_{\ell}^*(x)
\left[
    -
    \frac{\hbar^2}{2m}
    \frac{\partial^2}{\partial x^2}
    +
    V_{\rm L}(x)
\right]
W_{\ell+1}(x)
\,dx
\;,
\nonumber\\
U
&=&
g_{\rm 1D}
\int_{0}^{La_0}
\left|
    W_{\ell}(x)
\right|^4
\,dx
\;,
\end{eqnarray}
where $W_{\ell}(x)$ are the Wannier function for the lowest Bloch band (maximally) localized near the minima $x=x_\ell$
of the periodic potential $V_{\rm L}(x)$.
In the following, we obtain results for the Bose-Hubbard model by exact diagonalization, and for the  Hamiltonian in Eq.~(\ref{Eq:H:continuous}) with the DMC method.

\section{Excitation spectrum in the Tonks-Girardeau regime\label{Sec:TG}}

In the Tonks-Girardeau limit ($g_{\rm 1D} = +\infty$ or equivalently $a_{\rm 1D}= 0$), the wave function of $N$ bosons $\psi_{\rm B}$ can be mapped onto the wave function of $N$ non-interacting fermions
$\psi_{\rm F}$ as
\begin{equation}
\psi_{\rm B}(x_1,\dots,x_N)
=
\left|
    \psi_{\rm F}(x_1,\dots,x_N)
\right|
\;.
\end{equation}
This equality leads to identical ground-state energy and diagonal properties that depend on $|\psi|^2$.
In the language of second quantization, the Bose-Fermi mapping corresponds to the following transformation between the
bosonic and fermionic field operators (see Eq.~(16.75) in Ref.~\cite{Sachdev}):
\begin{eqnarray}
\hat\psi_{\rm B}(x)
=
\exp
\left[
    i\pi
    \int_0^x
    \hat\psi_{\rm F}^\dagger(x')
    \hat\psi_{\rm F}^{\phantom{\dagger}}(x')
    d\,x'
\right]
\hat\psi_{\rm F}(x)
\;.
\end{eqnarray}
We then impose periodic boundary conditions for bosons, $\hat\psi_{\rm B}(La_0)=\hat\psi_{\rm B}(0)$, which is equivalent to the requirement
\begin{equation}
\label{psiF}
\hat\psi_{\rm F}(La_0)
=
\exp
\left(
    -i\pi\hat N
\right)
\hat\psi_{\rm F}(0)
\end{equation}
for fermions.
Since the field operator can be represented as a superposition of single-particle modes
\begin{equation}
\hat\psi_{\rm F}(x)
=
\sum_{k}
\varphi_k(x)
\hat c_k
\;,
\end{equation}
Eq.~(\ref{psiF}) implies that
\begin{equation}
\varphi_k(La_0)=(-1)^{N+1}\varphi_k(0)
\;,
\end{equation}
i.e., boundary conditions for fermions are periodic for odd $N$ and antiperiodic for even $N$.

The eigenstates of the Tonks-Girardeau gas are Slater determinants built from the single-particle eigenfunctions $\varphi_{k}(x)$ of energies $\varepsilon_{k}$.
According to the Bloch theorem, the solutions have the following form
\begin{equation}
\label{psib}
\varphi_k(x)
\equiv
\varphi_b(x;k)
=
u_b(x;k)
e^{ikx}
\;,\quad
\varepsilon_k
\equiv
\varepsilon_b(k)
\;,
\end{equation}
where $u_b(x;k)$ is a periodic function of $x$ with period $a_0$, and $b=0,1,\dots$ is the band index.
The wavenumber $k$ takes discrete values
\begin{eqnarray}
k_q^{\rm pbc}
=
\frac{\pi}{a_0}
\frac{2q}{L}
\;,\quad
k_q^{\rm abc}
=
\frac{\pi}{a_0}
\frac{2q+1}{L}
\;,
\end{eqnarray}
for periodic and antiperiodic boundary conditions, respectively, with $q$ being a positive or negative integer.
The eigenvalues $\varepsilon_b(k)$ and the eigenfunctions $\varphi_b(x;k)$ are periodic functions of $k$ with the period equal to the vector of the reciprocal lattice $k_L=2\pi/a_0$.
Then the single-particle eigenstates obtained for the same Bloch band $b$ but for different Brillouin zones,
labeled by $n=1,2,\dots$ and determined as intervals
\begin{eqnarray}
\frac{ka_0}{\pi}
\in
\left[
    -n, -n+1
\right]
\cup
\left[
    n-1, n
\right)
\;,
\end{eqnarray}
are equivalent to each other.
In order to deal with distinct solutions, one can consider
(i) all Bloch bands within the first Brillouin zone, or
(ii) the $b$-th Bloch band within the $(b+1)$-th Brillouin zone.
In order to keep the analogy with the homogeneous space, where no restrictions on the values of momentum are imposed, we use the second option.

In the ground state, particles occupy the $N$ lowest-energy single-particle states with momenta
\begin{equation}
\label{kq}
k_q
=
-k_{\rm F}
+
\frac{\pi}{a_0}
\frac{2q+1}{L}
\;,\quad
q=0,\dots,N-1
\;,
\end{equation}
where $k_{\rm F}=\pi N/(La_0)$ is the Fermi momentum, i.e., the radius of the ``Fermi sphere'' in the thermodynamic limit.
The Fermi momentum $k_{\rm F}$ and the vector of the reciprocal lattice $k_L$ are related to each other as $k_{\rm F}=f k_L/2$, where $f=N/L$ is the filling factor.
The energy of the ground state is given by
\begin{equation}
\label{E0}
E_0
=
\sum_{q=0}^{N-1}
\varepsilon(k_q)
\end{equation}
and the total momentum $\sum_{q=0}^{N-1}k_q=0$.
Note that Eqs.~(\ref{kq}),~(\ref{E0}) are valid both for even and odd $N$.

The excited states of the Tonks-Girardeau gas are obtained when one or more particles are promoted outside of the Fermi sphere.
The excitation energy vanishes at a special value of the momentum, $k=2k_{\rm F}$, corresponding to the {\it umklapp} process,
in which a particle is moved from one side of the Fermi surface to the other side.
This creates an excitation of finite momentum but zero energy.
Physically, it is important to locate the position of the lowest branch $E_1(k)$ of the excitation spectrum.
We start with an overview of how excitations are generated in a homogeneous Tonks-Girardeau gas and then consider an optical lattice.

\subsection{Homogeneous space\label{Sec:TG:hom}}

In the homogeneous space case ($V_0=0$), the single-particle eigenstates are plane waves
\begin{equation}
\label{eikx}
\varphi_{k}^{\rm hom}(x)
=
\frac{1}{\sqrt{La_0}}
\exp
\left(
    i k x
\right)
\;.
\end{equation}
The momentum $k$ takes discrete values and the single-particle energies are given by $\varepsilon_k = \hbar^2k^2/(2m)$.

The highest level occupied by the particles in the many-body ground state has an energy
\begin{equation}
\varepsilon_{\rm max}
=
\varepsilon(k_0)
=
\varepsilon_{\rm F}
\left(
    \frac{N-1}{N}
\right)^2
\;,\quad
\varepsilon_{\rm F}
=
\frac{\hbar^2 k_{\rm F}^2}{2m}
\;,
\end{equation}
while the total energy of the ground state is obtained from the summation in Eq.~(\ref{E0}), and is given by
\begin{equation}
E_0^{\rm hom}
=
\frac{N}{3}
\varepsilon_{\rm F}
\left(
    1 - \frac{1}{N^2}
\right)
\;.
\end{equation}

\begin{figure}
\includegraphics*[width=\columnwidth,angle=0]{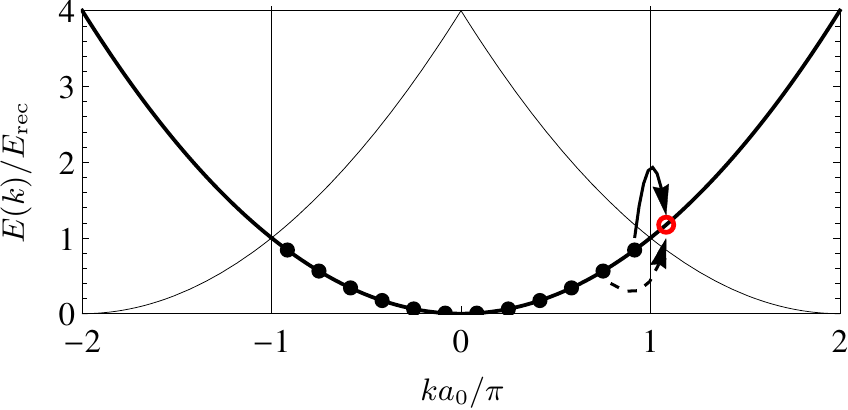}
\caption{(color online)
Occupation of the single-particle energy levels in a homogeneous space ($V_0=0$) for the ground state and low-energy excited states.
Black solid circles show the ground-state configuration for $N=L=12$.
The lowest-energy excitation branch~(\ref{Eq:E(k)homIFG}) is obtained by removing a particle from one of the occupied states and putting it following the solid arrow into the empty state marked by the open red circle.
The subsequent excitation process is obtained following the dashed arrow.
}
\label{FigExcHom}
\end{figure}

The excitations are created by promoting one or several particles from the ground state to the higher single-particle levels (single or multiple ``particle-hole'' excitations).
The lowest-energy excited state with a given momentum corresponds to a single-particle excitation, and is constructed by moving a particle from an occupied single-particle state to the empty single-particle state with $q=-1$ or $q=N$ in Eq.~(\ref{kq}) (Fermi surface), see Fig.~\ref{FigExcHom}.
These ``hole" excitations possess a momentum $k\equiv k_q=2\pi q/(La_0)$ and form a branch
\begin{equation}
\label{Eq:E(k)homIFG}
E_1^{\rm hom}(k_q)
=
\frac{\hbar^2}{2m}
\left[
    k_{\rm F}^2
    -
    \left(
        k_{\rm F} - k_q
    \right)^2
    +
    \frac{2}{N}
    k_{\rm F} k_q
\right]
\;,
\end{equation}
where the last term describes the finite-size effects.
Thus, the lowest-lying excitation in a homogeneous Tonks-Girardeau gas has a linear (phononic) spectrum,
$E_1^{\rm hom}(k) = \hbar c |k|$ at $k \to 0$, corresponding to a sound velocity $c=\hbar k_{\rm F}/m$.

\subsection{Optical lattice\label{Sec:TG:lattice}}
\begin{figure}
\includegraphics*[width=\columnwidth,angle=0]{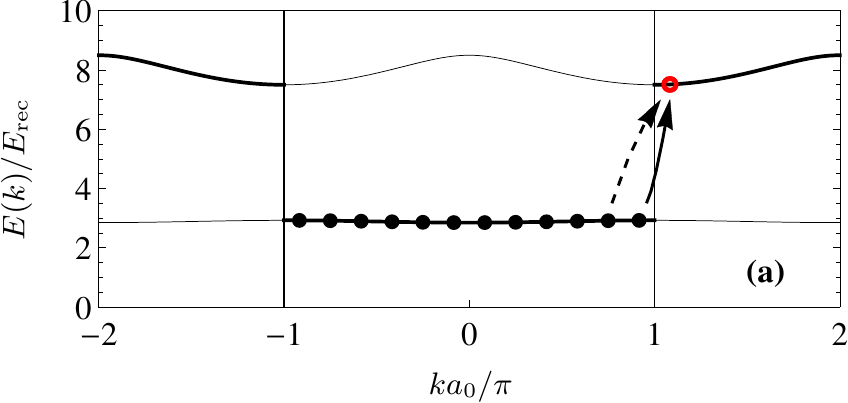}
\includegraphics*[width=\columnwidth,angle=0]{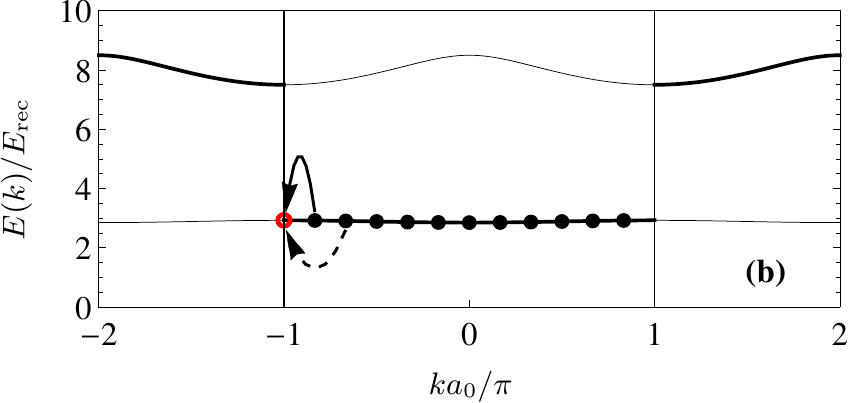}
\includegraphics*[width=\columnwidth,angle=0]{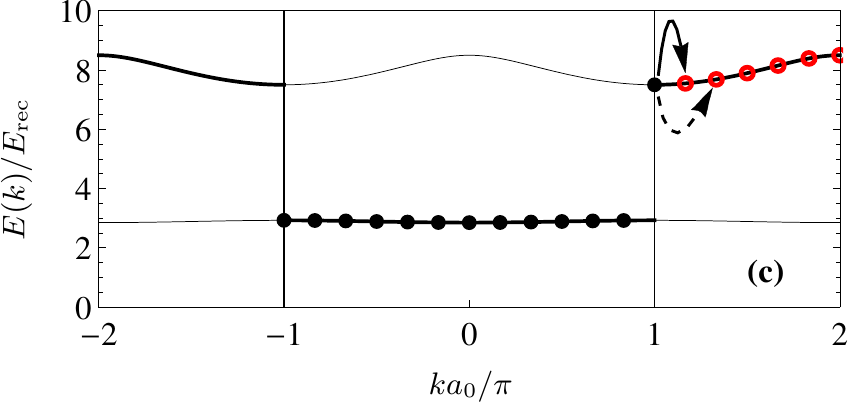}
\caption{(color online)
Occupation of the single-particle energy levels in a deep lattice ($V_0 = 10\, E_{\rm rec}$) of length $L=12$ for the ground state and low-energy excited states.
The following numbers of particles are used
(a) $N=L$ (commensurate, unit filling),
(b) $N=L-1$ (incommensurate, one vacancy),
(c) $N=L+1$ (incommensurate, one interstitial).
Black solid circles correspond to the ground-state configurations.
Arrows and red open circles indicate processes that lead to the lowest-energy excitations with different momenta.
}
\label{FigExcLat}
\end{figure}

\begin{figure}

\includegraphics*[width=0.8\columnwidth,angle=0]{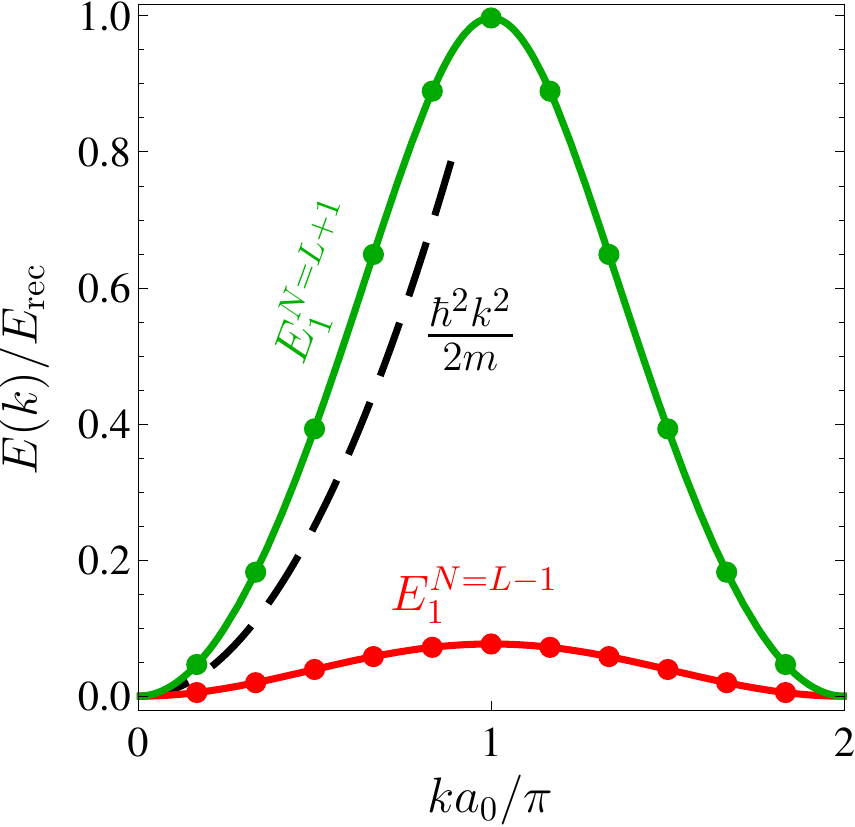}
\caption{
(color online)
Energy of the lowest excitations in a deep lattice ($V_0 = 10\,E_{\rm rec}$)
for $N=L-1$~(red), $N=L+1$~(green).
Solid circles: discrete allowed values of the momentum for system size $L=12$;
solid lines: continuous values of the momentum in the limit of infinitely-large lattice ($L\to\infty$).
The excitation branch $E_1^{N=L}(k)$ has the same form as $E_1^{N=L-1}(k)$ but shifted upwards by the energy gap $\Delta$.
The corresponding effective masses defined by Eqs.~(\ref{E1NL}),~(\ref{E1NL+1}) are
$m_{\rm eff} = 5.03\,m$ (vacancy) and $m_{\rm int} = 0.59\,m$ (interstitial).
Dashed line corresponds to the homogeneous space ($V_0=0$) with the energy of free particles $\hbar^2k^2/(2m)$.
}
\label{Fig:BZ}
\end{figure}

When $V_0>0$ the excitation spectrum of the Tonks-Girardeau gas strongly depends on the ratio between $N$ and $L$.
Here we consider three cases: $N=L$, $N=L-1$ and $N=L+1$~(Fig.~\ref{FigExcLat}), corresponding to a unit filling, one vacancy and one interstitial, respectively.

In the case of unit filling ($N=L$, Fig.~\ref{FigExcLat}a), the ground state corresponds to the full population of the first Brillouin zone.
As in the homogeneous case, the lowest excitations with different momenta $k_q$ are created by removing a particle from one of the occupied states and inserting it into the state $k_{-1}$ or $k_N$. The corresponding branch of excitations is given by
\begin{equation}
E_1^{N=L}(k_q)
=
\varepsilon(k_N)
-
\varepsilon(k_{N-q})
\;,\quad
q=0,\dots,N-1
\;.
\end{equation}
As the states with $k_{-1}$ and $k_N$ belong to the second Brillouin zone, there is a gap  $\Delta=\varepsilon(k_N)-\varepsilon(k_{N-1})$ in the excitation spectrum.
In a finite-size system the momenta are discretized as shown in Fig.~\ref{FigExcLat}.
As the system size is increased the discretization becomes smaller, but the form the single-particle levels remains similar and the gap survives in the thermodynamic limit.
For small $k$, the spectrum can be described by the value of the gap $\Delta$ plus a quadratic dispersion relation, which defines the effective mass $m_{\rm eff}$ according to
\begin{equation}
\label{E1NL}
E_1^{N=L}(k)
\approx
\Delta
+
\frac{\hbar^2 k^2}{2m_{\rm eff}}
\;.
\end{equation}
This situation is entirely different from the gapless excitation of the homogeneous gas, Eq.~(\ref{Eq:E(k)homIFG}),
which presents a linear dependence on momentum at low $k$.

In the presence of a single vacancy ($N=L-1$, Fig.~\ref{FigExcLat}b), the ground-state configuration contains one empty single-particle state at the edge of the first Brillouin zone and the lowest excitations are created by promoting a particle into this state.
This mechanism removes the gap from the excitation spectrum in the thermodynamic limit ($N=L-1$, $L\to\infty$) but leaves the same $k$-dependence as in the $N=L$ case discussed above.
The lowest excitation branch $E_1^{N=L-1}(k)$ is shown in Fig.~\ref{Fig:BZ}.

This quadratic dependence applies only to the case of a microscopic number of vacancies $N_{\rm vac}$ ($N_{\rm vac}$ much smaller than the total number of atoms $N$).
Instead, when a macroscopic fraction of vacancies is introduced ($N_{\rm vac}$ proportional to $N$), the excitation spectrum becomes linear.
Importantly, the common feature of introducing either a microscopic or a macroscopic number of vacancies is that in the both cases the excitation spectrum changes from gapped to a gapless one.
To a certain extent, the mechanism is similar to that found in semiconductors, where an insulator can be transformed into a conductor by a doping which injects holes or electrons into a fully-filled crystal structure.

In the presence of an interstitial ($N=L+1$, Fig.~\ref{FigExcLat}c), the ground-state configuration consists of a fully occupied first Brillouin zone plus another particle in one state in the second Brillouin zone.
The lowest excitations are created by promoting a particle into empty states of the second Brillouin zone.
The corresponding excitation branch is shown in Fig.~\ref{Fig:BZ}.
For small $k$, it has the form of a free particle spectrum
\begin{equation}
\label{E1NL+1}
E_1^{N=L+1}(k)
\approx
\frac{\hbar^2 k^2}{2m_{\rm int}}
\;,
\end{equation}
with an effective mass $m_{\rm int}<m$, see Fig.~\ref{Fig:BZ}.
In the particular case of a deep optical lattice with $V_0 = 10 E_{\rm rec}$, the effective mass of a vacancy is $m_{\rm eff} = 5.03\,m$ and that of an interstitial is $m_{\rm int} = 0.59\,m$.
These predictions, obtained from the low-energy expansion of the energetic spectrum, will be confronted to the estimation of the effective mass from the diffusion coefficient in Section~\ref{Sec:finite interaction}.

\begin{figure}
\includegraphics*[width=\columnwidth,angle=0]{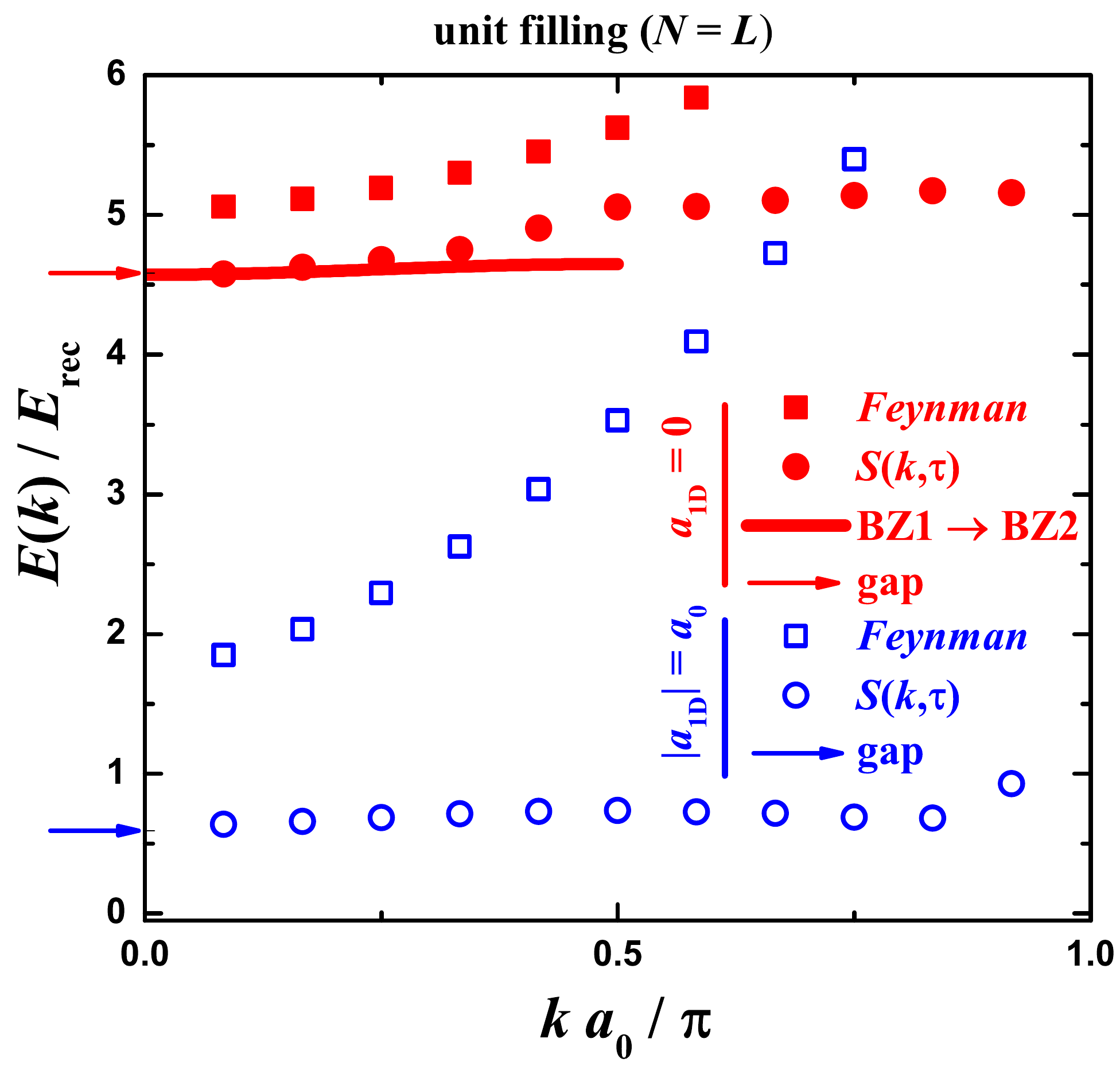}
\caption{(color online)
Lowest branch of the excitation spectrum $E(k)$ for unit filling ($N=L=12$) in a deep lattice with $V_0 = 10\,E_{\rm rec}$.
Solid symbols: $s$-wave scattering length $a_{\rm 1D}=0$ (Tonks-Girardeau limit);
open symbols: $|a_{\rm 1D}| = a_0$.
Solid line: the exact position for the transitions from the first into the second Brillouin zone for $a_{\rm 1D}=0$).
Squares: upper bound provided by Feynman relation~(\ref{Eq:Feynman}) from the static structure factor $S(k)$, mixed estimator.
Circles: single-exponent fit~(\ref{Eq:gap:tau}) to the long-imaginary time asymptotics of the dynamic structure factor.
Arrows show the value of the gap, obtained as the chemical potential difference~(\ref{Eq:gap:E}).
}
\label{Fig:Feynman:comensurate}
\end{figure}

\begin{figure}
\includegraphics*[width=\columnwidth,angle=0]{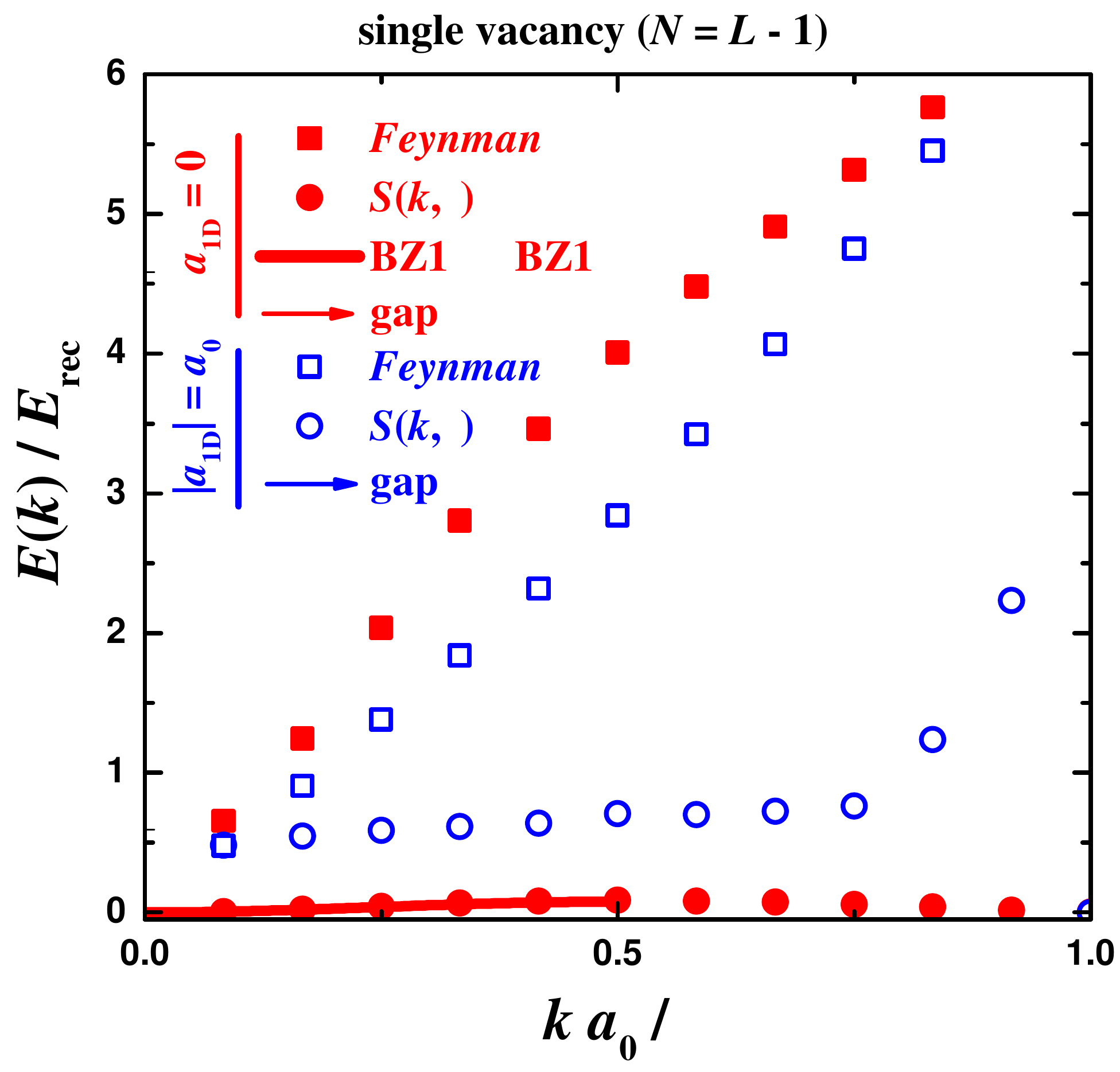}
\caption{(color online)
Same as in Fig.~\ref{Fig:Feynman:comensurate} but for an incommensurate filling (single vacancy with $N=L-1$ particles).
Solid line, the exact position for the transitions within the first Brillouin zone for $a_{\rm 1D}=0$).
}
\label{Fig:Feynman}
\end{figure}

\section{Excitation spectrum for finite interaction\label{Sec:finite interaction}}

Once reviewed the characteristic properties of the excitation spectrum of the infinitely-repulsive Tonks-Girardeau (``fermionic' ') gas, we consider the case of a finite interaction strength (``bosonic'' gas).
In this case, the Bose-Fermi mapping no longer applies, and the problem of finding the excitation spectrum of the quantum many-body system becomes a highly nontrivial task.
Fortunately, the diffusion Monte Carlo method gives access to imaginary-time dynamics and permits to extract the position of the lower branch of the excitation spectrum $\omega(k)$ by fitting the dynamic structure factor to an exponentially decaying law
\begin{equation}
S(k,\tau) \propto \exp(-\omega(k)\tau)\;,
\label{Eq:gap:tau}
\end{equation}
for asymptotically large imaginary times $\tau$.
We test this approach in the infinitely repulsive case, and compare with the exact analysis in terms of single-particle excitations (see Figs.~\ref{Fig:Feynman:comensurate}-\ref{Fig:Feynman}).
Having found a reasonable agreement, we apply the imaginary-time method to a system with finite interaction strength, taking $|a_{\rm 1D}| / a_0 = 1$ as a characteristic example.
As expected~\cite{Astrakharchik16}, at unit filling the system is insulating and a gapped excitation spectrum is observed (refer  to Fig.~\ref{Fig:Feynman}).
The value of the gap can be extracted in an alternative way from the difference of chemical potentials between the $N\pm1$ and the $N$ particle systems
\begin{equation}
\Delta = E_{N+1} - 2 E_N + E_{N-1}
\;,
\label{Eq:gap:E}
\end{equation}
which is shown in Fig.~\ref{Fig:Feynman:comensurate} as arrows at $k=0$.
We find a good agreement between the imaginary-time method~(\ref{Eq:gap:tau}) and the ground-state difference~(\ref{Eq:gap:E}).
In the limit of strong interactions and shallow lattices, the gap scales linearly with the lattice height, $\Delta \approx V_0/2$.
For finite interaction the gap becomes smaller and no analytical expression is known.
In the insulating phase the value of the gap is reduced by increasing the coupling constant.
Close to the phase transition point the gap becomes non-analytic, as it is zero in the superfluid phase and takes finite value the Mott-insulator one, and thus it cannot admit a Taylor expansion at the transition point.
Since the transition is of the Berezinsky-Kosterlitz-Thouless type, the gap closes as
\begin{eqnarray}
\Delta
\sim
\exp
\left(
    -\frac{\rm const}{\sqrt{|a_{\rm 1D}-a_{\rm c}|}}
\right)
\;,
\end{eqnarray}
where $a_{\rm c}$ is the critical value of the $s$-wave scattering length.

It is instructive to compare the predictions for the excitation spectrum $E(k)$ obtained from the static structure factor
$S(k)$ according to the Feynman relation
\begin{eqnarray}
E(k) \leq \frac{\hbar^2k^2}{2mS(k)}\;,
\label{Eq:Feynman}
\end{eqnarray}
which provides an upper bound to $E(k)$.
The Feynman approximation becomes exact when the excitation spectrum is exhausted by a single excitation.
We test the accuracy of Eq.~(\ref{Eq:Feynman}) by comparing it to the exact results, as shown in Fig.~\ref{Fig:Feynman:comensurate}.
We find out that for the gapped excitation at unit filling, the Feynman approximation works reasonably well.
This suggests that there is a strong weight of excitations at the gap and its value should be accessible experimentally,
on which we will comment in more details below.

Instead, for a single vacancy the Feynman approximation~(\ref{Eq:Feynman}) predicts a linear slope of the spectrum for small momenta, while the exact result shows a quadratic dependence, see Fig.~\ref{Fig:Feynman}.
On one hand, there is no contradiction, as the Feynman approximation provides an upper bound.
On the other hand, such difference implies that the quadratic gapless mode is not highly populated and it will be hard to see it in an experiment.
In all cases the Feynman approximation fails at the edge of the Brillouin zone $k=k_{L}/2$, where the correct excitation spectrum must have a zero derivative, and cannot reproduce the zero energy at the {\it umklapp} point $k=2k_{\rm F}$.

The superfluid -- Mott-insulator phase transition in a one-dimensional optical lattice was experimentally observed in the Innsbruck~\cite{Haller10} and Florence~\cite{Boeris15} groups.
Unfortunately, only in the Innsbruck experiment the excitation spectrum was measured.
The value of the gap was determined by shaking the lattice and analyzing the efficiency of the energy transfer for different driving frequencies.
The inferred experimental values of the gap turned out to be large compared to a subsequent theoretical  prediction~\cite{Astrakharchik16}.
This discrepancy leaves the question of the excitation spectrum still open.

\section{Superfluid fraction\label{Sec:sf}}

The superfluidity can be quantified by imposing twisted boundary conditions on the many-body  wavefunction~\cite{FBJ73,SS90,KrauthTrivediCeperley91}
\begin{equation}
\Psi(\dots,{x}_j+La_0,\dots)
=
e^{i\theta}
\Psi(\dots,{x}_j,\dots)
\;,
\end{equation}
$\theta$ being the twist angle.
At $T=0$, this requirement leads to an increase of the ground-state energy $E_0$, which is attributed to the kinetic energy of the superfluid.
Since the corresponding velocity is fixed by the value of the twist angle $\theta$, the number of particles in the superfluid component $N_{\rm s}$ is determined as
\begin{equation}
\label{Ns-def}
E_0(k_{\rm s})
-
E_0(0)
=
\frac{\hbar^2 k_{\rm s}^2}{2m}
N_{\rm s}
\;,\quad
k_{\rm s}
=
\frac{\theta}{La_0}
\;,
\end{equation}
which readily gives the superfluid fraction $N_{\rm s}/N\in[0,1]$.
In the limit $k_{\rm s}\to0$, Eq.~(\ref{Ns-def}) yields
\begin{equation}
\label{nus}
\label{Eq:sf:phasetwist}
N_{\rm s}
=
\frac{m}{\hbar^2}
\left.
    \frac{\partial^2 E_0(k_{\rm s})}{\partial k_{\rm s}^2}
\right|_{k_{\rm s}=0}
\;,
\end{equation}
and, for instance, in the case of noninteracting particles in a periodic potential, where $E_0(k_{\rm s}) = N \varepsilon(k_{\rm s})$, one gets $N_{\rm s}/N=m/m_{\rm eff}$~\cite{Eggington77,AHNS80}.

The superfluid fraction $N_{\rm s}/N$ of the Tonks-Girardeau gas in deep optical lattices
for $N\leq L$ can be calculated analytically and is given by~\cite{K16}
\begin{equation}
\label{fshchom}
\frac{N_{\rm s}}{N}
=
\frac{m}{m_{\rm eff}}
\frac{\sin (\pi N/L)}{N\sin(\pi/L)}
\;,
\label{Eq:sf:TG}
\end{equation}
where $m_{\rm eff}$ is the single-particle effective mass.
For a single vacancy, $N=L-1$, Eq.~(\ref{Eq:sf:TG}) simplifies to
\begin{eqnarray}
\frac{N_{\rm s}}{N}
=
\frac{m}{m_{\rm eff}}
\frac{1}{N}
\;.
\label{Eq:sf:vacancy}
\end{eqnarray}
Similarly to the Anreev-Lifshitz mechanism in which vacancies in the ground-state of a solid create superfluidity, here the presence of vacancies in the lattice lead to a non-zero value of $N_{\rm s}$.
The total effect is microscopic since the contribution of a single vacancy to the superfluid fraction is of order $1/N$.
Equation~(\ref{Eq:sf:vacancy}) implies that a single vacancy turns the system completely superfluid, while its contribution to the superfluid fraction is reduced by the effective mass.

While Eq.~(\ref{Eq:sf:TG}) was derived only for vacancies and it is not formally applicable to interstitials, still for a single interstitial Eq.~(\ref{Eq:sf:TG}) with $N = L + 1$ leads to a result similar to Eq.~(\ref{Eq:sf:vacancy}) but of opposite sign.
Thus it is a surprise to find out that numerical simulations based on the diffusion Monte Carlo method confirm that Eq.~(\ref{Eq:sf:vacancy}) holds even for interstitials, as it will be shown later.

\begin{figure}
\includegraphics[width=\columnwidth,angle=0]{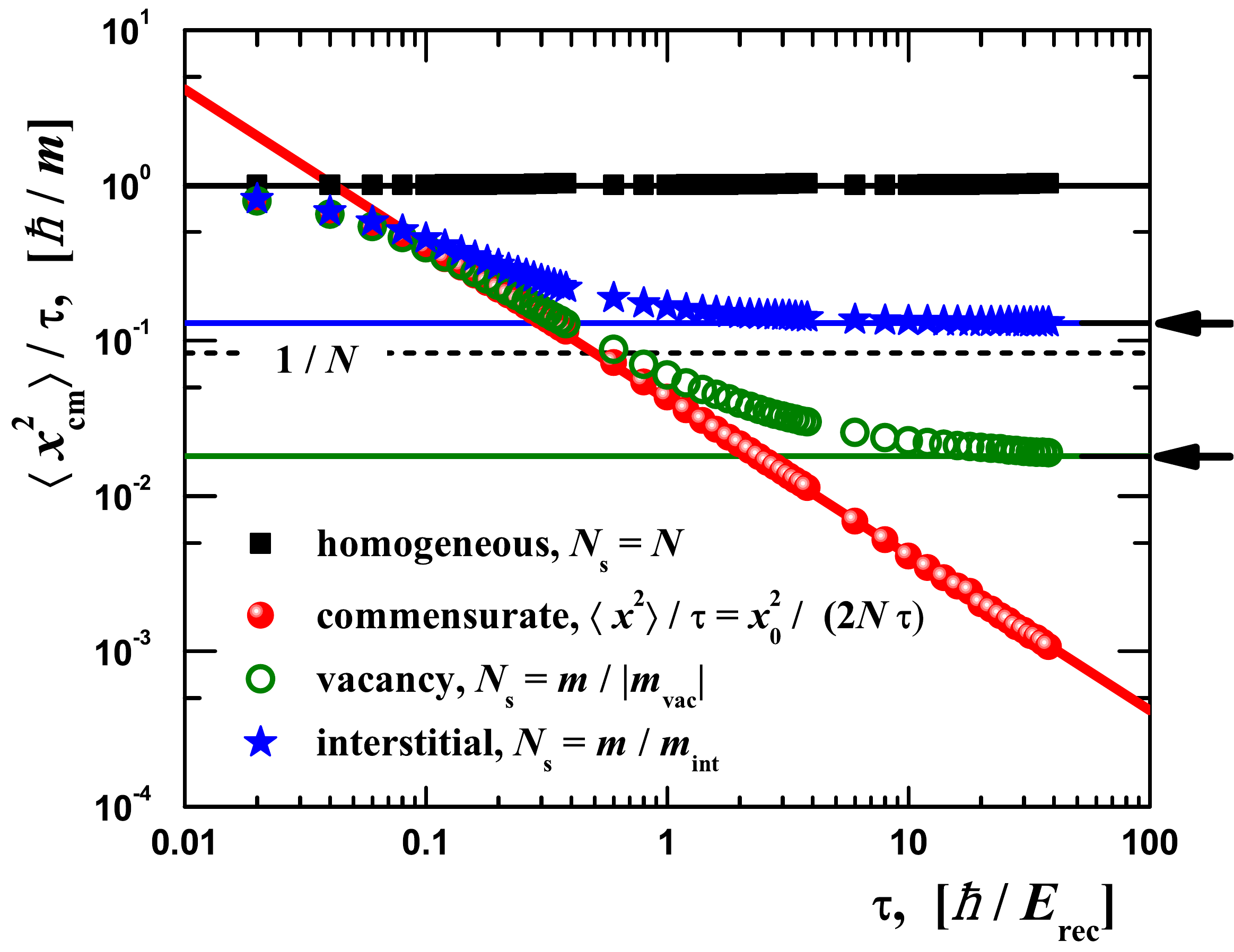}
\caption{(color online)
Diffusion of the center of mass in imaginary time,
$\langle x^2(\tau)\rangle /\tau$,
for Tonks-Girardeau gas in a deep lattice, $V_0 = 10\,E_{\rm rec}$.
A constant value at asymptotically large imaginary times gives the superfluid fraction $N_{\rm s}/N$.
Symbols, DMC data;
two arrows on the right axis, energetic estimation from a small phase twist, Eq.~(\ref{Eq:sf:phasetwist});
lines:
homogeneous system, $N_{\rm s} = N$ (arbitrary filling);
commensurate system, $\langle x^2(\tau)\rangle = 1/(2N\tau)$;
vacancy, Eq.~(\ref{Eq:sf:vacancy}) with $m_{\rm vac} = m_{\rm eff} = 5.03\,m$,
interstitial, Eq.~(\ref{Eq:sf:vacancy}) with $m_{\rm int} = 0.59\,m$.
Dashed line with $N_{\rm s} = 1$ is shown for comparison and allows us to distinguish a defect with the effective mass smaller or larger than one.
At very short imaginary times, the diffusion is always ballistic and the shown quantity always departs from one.
}
\label{Fig:SD}
\end{figure}

Applied to the DMC algorithm, the {\it winding number} method for the estimation of the superfluid density~\cite{Ceperley95}, Eq.~(\ref{Eq:sf:phasetwist}), is equivalent to the calculation of the diffusion coefficient of the center of mass.
The center of mass position is known from the particle position, $x_{\rm cm} = \sum_{i=1}^N x_i / N$, and we measure the mean square distance $\langle x_{\rm cm}^2 \rangle = \langle [x_{\rm cm}(\tau) -x_{\rm cm}(0)]^2 \rangle$ it diffuses in a time $\tau$.
Characteristic examples of the diffusion of the center of mass in the Tonks-Girardeau gas are shown in Fig.~\ref{Fig:SD}.
In the absence of lattice, the superfluid response is macroscopic with $N_{\rm s} = N$ for any filling fraction.
This could be commonly interpreted as a fully superfluid homogeneous system, although in one dimension this is pathological, as even a tiny external potential greatly changes the superfluid response.
Indeed, the Tonks-Girardeau gas can be mapped to an ideal Fermi gas, which is not superfluid and suffers from the {\it orthogonality catastrophe}, where the wave function in the presence of a tiny perturbation is asymptotically orthogonal to the wave function of the unperturbed state.
In lattice systems a similar effect in commensurate systems is present not only for infinite but also for finite interactions, for which the Mott insulator is formed for arbitrarily small height of the lattice.

This effect is also seen in a Tonks-Girardeau gas in a commensurate lattice as the system is no longer superfluid.
Indeed, a direct energy calculation of the Tonks-Girardeau energy plugged into Eq.~(\ref{Eq:sf:phasetwist}) gives zero superfluid fraction.
In DMC simulations, we find that the center of mass is pinned and its spreading $\langle x_{\rm cm}^2\rangle$ is bounded from above for large times, resulting in $\langle x_{\rm cm}^2\rangle / \tau \propto 1/\tau \to 0$ for $\tau\to\infty$.

The presence of a single vacancy enables defect-induced superfluidity, similarly to the Andreev-Lifshitz mechanism.
The superfluid response is microscopic, $N_{\rm s} = m/m_{\rm eff} = O(1)$, as only one defect contributes to the superfluidity.
Furthermore, the contribution to the superfluid component is further reduced by the effective mass.
The propagation of a vacancy physically corresponds to a quasi-particle created by atoms tunneling from one site to another, with the quasi-particle propagating in the opposite direction.
It is natural that the movement of particles over a barrier is slower compared to the movement of a free particle, and so the effective mass is increased, $m_{\rm vac}>m$.
Specifically, for the considered case of a deep optical lattice $V_0 = 10\,E_{\rm rec}$, the vacancy is rather heavy,
with $m_{\rm vac} = 5.03\,m$, see Fig.~\ref{Fig:BZ}.

A single interstitial introduces a finite superfluid response as well.
The contribution is again microscopic and is scaled by the effective mass of the defect.
Remarkably, Eq.~(\ref{Eq:sf:vacancy}) with $m_{\rm int}$ taken from the quadratic expansion around the edge of the second Brillouin zone gives exactly the same result as the energetic estimation~(\ref{Eq:sf:phasetwist}) and agrees with the diffusion of the center-of-mass method in the DMC calculation.
This verifies the applicability of Eq.~(\ref{Eq:sf:vacancy}) and also validates the microscopic description of the superlfuidity in terms of defects.

\section{Static structure factor\label{Sec:Sk}}

Relevant information about the spatial structure of the system is contained in the static structure factor.
In the superfluid phase, the long-range (small-momenta) correlations are well captured by the Luttinger liquid theory.
Both for bosons and fermions, the characteristic length is the mean interparticle distance $La_0/N$, or the Fermi momentum $k_{\rm F} = \pi N/(La_0)$ in the reciprocal space.
The Luttinger liquid theory predicts oscillations at multiples of the Fermi momentum, $k=k_{\rm F}, 2k_{\rm F}, \dots$.
The lattice provides an additional, external scale, which in terms of length and momentum is the lattice spacing $a_0$
and the edge of the Brillouin zone $k_{L}/2$, respectively.
At unit filling both scales coincide, $k_{\rm F} = k_{L}/2$ while for a fractional filling the Fermi momentum can be expressed as $k_{\rm F} = f k_{L}/2$.
It can be anticipated that the correlation functions at momentum $k$, corresponding to the difference
$k=k_{L}/2-k_{\rm F} = (1-f)k_{L}/2$, have special properties.
In the following we analyze the cases of both a close mismatch between $k_{L}/2$ and $k_{\rm F}$ (i.e., in the presence of few vacancies or interstitials) as well as a large one (macroscopic fraction of vacancies) showing that elaborate correlations appear in the system.

\subsection{Continuum model}

In order to quantify the effect of correlations, we consider the static structure factor which has two commonly used definitions.
The first one is given by
\begin{eqnarray}
\label{Srho}
S_\rho({k})
&=&
\frac{1}{N}
\int_0^{La_0} \!\! d{x}_1
\int_0^{La_0} \!\! d{x}_2
e^{
   i{k}
   \left(
       {x}_2 - {x}_1
   \right)
  }
\langle
    \hat\rho({x}_1)
    \hat\rho({x}_2)
\rangle
\\
&=&
1
+
\frac{1}{N}
\int_0^{La_0} \!\! d{x}_1
\int_0^{La_0} \!\! d{x}_2
e^{
   i{k}
   \left(
       {x}_2 - {x}_1
   \right)
  }
g_2(x_1,x_2)
\;,
\nonumber
\end{eqnarray}
where $\hat\rho({x})=\hat\psi^\dagger({x})\hat\psi({x})$ is the density operator.
It is closely related to the Fourier transform of the pair distribution function
that can be written in the language of second and first quantization as
\begin{eqnarray}
&&
g_2(x_1,x_2)
=
\langle
   \hat\psi^\dagger({x}_1)
   \hat\psi^\dagger({x}_2)
   \hat\psi({x}_2)
   \hat\psi({x}_1)
\rangle
\\
&&
=
N(N-1)
\int_0^{La_0}
\!\!
\left|
    \psi({x}_1,{x}_2,{x}_3,\dots,{x}_N)
\right|^2
d{x}_3 \dots d{x}_N
\;.
\nonumber
\end{eqnarray}
The second definition uses density fluctuations
$\Delta\hat\rho(x)=\hat\rho(x)-\langle\hat\rho(x)\rangle$ instead of the density $\hat\rho(x)$ and has the explicit form
\begin{eqnarray}
S_{\Delta\rho}({k})
=
\frac{1}{N}
\int_0^{La_0} \!\! d{x}_1
\int_0^{La_0} \!\! d{x}_2
e^{
   i{k}
   \left(
       {x}_2 - {x}_1
   \right)
  }
\langle
    \Delta\hat\rho({x}_1)
    \Delta\hat\rho({x}_2)
\rangle
\;,
\label{Sdrho}
\end{eqnarray}
while the two definitions are related to each other as
\begin{eqnarray}
S_{\rho}({k})
=
S_{\Delta\rho}({k})
+
\frac{1}{N}
\left|
    \int_0^{La_0} dx
    e^{-ikx}
    \langle
        \hat\rho(x)
    \rangle
\right|^2
\;.
\label{SrhoSdrho}
\end{eqnarray}

As we will show later in the figures, $S_{\rho}({k})$ contains singularities at $k=qk_L$ with integer $q$ originating from the last term in Eq.~(\ref{SrhoSdrho}) and induced by the lattice periodicity.
In order to illustrate this point, we expand the density profile in a Fourier series,
\begin{equation}
\langle
    \hat\rho(x)
\rangle
=
\sum_n
c_n
\exp
\left(
    i2\pi n \frac{x}{a_0}
\right)
\;,
\end{equation}
where $c_0=N/(La_0)$ is the averaged density and $c_{-n}=c_n^*$ since $\langle\hat\rho(x)\rangle$ is real.
The momentum $k$ takes discrete values and it is easy to see that the last term in Eq.~(\ref{SrhoSdrho}) is non-zero only for $k=qk_L$ with $q=0,\pm1,\cdots$.
In this way, we get
\begin{eqnarray}
\label{Srho_peak}
S_{\rho}(qk_L)
=
S_{\Delta\rho}(qk_L)
+
N
\left|
 \frac{c_q}{c_0}
\right|^2
\;,
\end{eqnarray}
and thus if $|c_q|$ is of order of unity, $S_{\rho}(k)$ presents strong peaks at $k=qk_L$ on top of $S_{\Delta\rho}(k)$ and the height of the peaks grows linearly with $N$.

For the Tonks-Girardeau gas, the pair distribution function is the same as for free fermions.
Therefore, it can be written in terms of single-particle eigenfunctions $\varphi_\alpha(x)$ as~\cite{LW07,WW10,BZS10}
\begin{eqnarray}
g_2(x_1,x_2)
=
\frac{1}{2}
\sum_{\alpha_1,\alpha_2=0}^{N-1}
\left|
    \varphi_{\alpha_1}(x_1)
    \varphi_{\alpha_2}(x_2)
    -
    \varphi_{\alpha_1}(x_2)
    \varphi_{\alpha_2}(x_1)
\right|^2
\;,
\nonumber
\end{eqnarray}
leading to the following expression for the static structure factor,
\begin{eqnarray}
\label{Sk}
&&
S_{\Delta\rho}(k)
=
1
-
\frac{1}{N}
\sum_{\alpha_1,\alpha_2=0}^{N-1}
\left|
    I_{\alpha_1 \alpha_2}(k)
\right|^2
\;,
\\
&&
I_{\alpha_1 \alpha_2}(k)
=
\int_{0}^{La_0}
dx
e^{-ikx}
\varphi_{\alpha_1}^*(x)
\varphi_{\alpha_2}(x)
\;.
\nonumber
\end{eqnarray}
In homogeneous space, the single-particle functions $\varphi^{\rm hom}_\alpha(x)$ are plane waves, Eq.~(\ref{eikx}), with the discrete values of momenta given by Eq.~(\ref{kq}). Then,
\begin{equation}
I^{\rm hom}_{q_1 q_2}(k_q)
=
\delta_{q_2-q_1,q}
\;,
\end{equation}
and Eq.~(\ref{Sk}) leads to the well-known result for a homogeneous system in one dimension,
\begin{equation}
S^{\rm hom}_{\Delta\rho}(k)
=
{\rm min}
\left(
    \frac{k}{2k_{\rm F}}\,,\,1
\right)
\;,
\quad
S^{\rm hom}_{\rho}(k)
=
S^{\rm hom}_{\Delta\rho}(k)
+
N\delta_{k0}
\;,
\label{Eq:Sk:TG:hom}
\end{equation}
which is a linear (phononic) up to the {\it umklapp} point $k=2k_F$, with a kink at this point, and followed by a constant plateau.

\subsection{Discrete model}

For discrete lattice models (such as the Bose-Hubbard model), one can define discrete structure factors in analogy to Eqs.~(\ref{Srho}),~(\ref{Sdrho}):
\begin{eqnarray}
S_{n}(k)
&=&
\frac{1}{N}
\sum_{\ell_1,\ell_2}
\langle
    \hat n_{\ell_1}
    \hat n_{\ell_2}
\rangle
e^{
    i k
    \left(
        x_{\ell_2}-
        x_{\ell_1}
    \right)
  }
\;,
\\
S_{\Delta n}(k)
&=&
\frac{1}{N}
\sum_{\ell_1,\ell_2}
\langle
    \Delta\hat n_{\ell_1}
    \Delta\hat n_{\ell_2}
\rangle
e^{
    i k
    \left(
        x_{\ell_2}-
        x_{\ell_1}
    \right)
  }
\;,
\end{eqnarray}
with $\Delta\hat n_{\ell}=\hat n_{\ell}-\langle\hat n_{\ell}\rangle$,
that are related to each other as
\begin{eqnarray}
S_{n}(k)
=
S_{\Delta n}(k)
+
\frac{1}{N}
\left|
    \sum_{\ell}
    e^{-ikx_{\ell}}
    \langle\hat n_\ell\rangle
\right|^2
\;.
\end{eqnarray}

In the tight-binding approximation described by the Bose-Hubbard model~(\ref{Eq:H:BH}), the static structure factors
of the continuum model are determined by the corresponding structure factors of the discrete model.
For $S_{\Delta\rho}(k)$ we get~\cite{K16}
\begin{eqnarray}
\label{S0lBB}
S_{\Delta\rho}^{\rm BH}(k)
=
1
+
G_0^2(k)
\left[
    S_{\Delta n}(k)
    -1
\right]
\;,
\end{eqnarray}
where
\begin{eqnarray}
G_0(k)
=
\int_0^{La_0}
d{x}
\left|
    W_\ell(x)
\right|^2
\exp
\left[
    -i k
    \left(
        x - x_\ell
    \right)
\right]
\;.
\end{eqnarray}
It is easy to see that $S_{\Delta n}(k)$ is a periodic function of $k$ with the period equal to the vector of the reciprocal lattice $k_L=2\pi/a_0$.
Also, $S_{\Delta n}(k)$ vanishes at special points $k=qk_L$, $q=0,1,\dots$, which implies that $S_{\Delta\rho}^{\rm BH}(qk_L)=1-G_0^2(qk_L)$ is determined entirely by the periodic potential and does not depend on the atomic interaction and filling factor.

\begin{figure}
\includegraphics*[width=\columnwidth,angle=0]{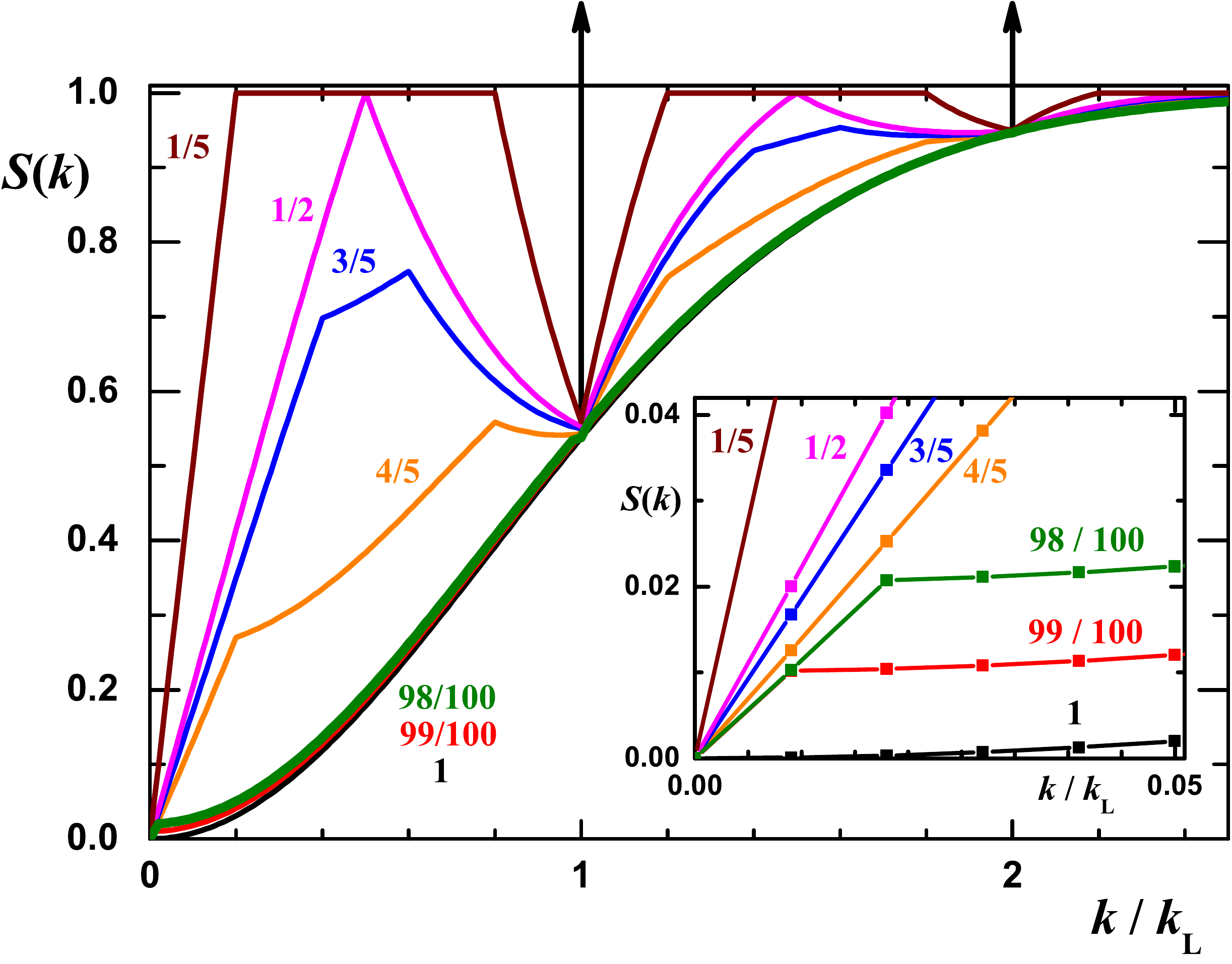}
\caption{(color online)
Static structure factor $S_{\rho}(k)$ of the Tonks-Girardeau gas in a deep lattice ($V_0 = 10\,E_{\rm rec}$) with $L=100$ lattice sites and fractional filling (decreasing the height of the peak) 1/5; 1/2; 3/5; 4/5; 99/100; 1.
The peaks at commensurate momentum $k = q k_L, q = 1,2,\cdots$ are macroscopically large and are denotes with arrows and the trivial peak $S_{\rho}(0)=N$ is not shown.
Inset: zoom for small momentum part.
}
\label{Fig:Sk:vac}
\end{figure}

For the Tonks-Girardeau gas in the tight-binding approximation with the filling $f\le1$ one has
\begin{eqnarray}
&&
\langle
    \hat n_{\ell_1}
    \hat n_{\ell_2}
\rangle
-
\langle
    \hat n_{\ell_1}
\rangle
\langle
    \hat n_{\ell_2}
\rangle
\\
&&
=
\langle
    \hat n_{\ell_1}
\rangle
\delta_{\ell_1\ell_2}
-
\left|
    \sum_{\alpha=0}^{N-1}
    \phi_{\alpha\ell_1}^*
    \phi_{\alpha\ell_2}
\right|^2
\nonumber
\end{eqnarray}
where $\phi_{\alpha\ell}$ are the single-particle eigenfunctions on a lattice.
The discrete structure factor can be rewritten in a form analogous to Eq.~(\ref{Sk}):
\begin{eqnarray}
\label{S0k}
&&
S_{\Delta n}(k)
=
1
-
\frac{1}{N}
\sum_{\alpha_1,\alpha_2=0}^{N-1}
\left|
    {\cal I}_{\alpha_1 \alpha_2}(k)
\right|^2
\;,
\\
&&
{\cal I}_{\alpha_1 \alpha_2}(k)
=
\sum_{\ell}
e^{-ikx_\ell}
\phi_{\alpha_1\ell}^*
\phi_{\alpha_2\ell}
\;.
\nonumber
\end{eqnarray}
Since the parameters of the Bose-Hubbard model do not depend on the lattice-site index,
\begin{equation}
\phi_{q\ell}
=
\frac{1}{\sqrt{L}}
\exp
\left(
     i k_q x_\ell
\right)
\;,
\end{equation}
where $k_q$ are determined by Eq.~(\ref{kq}), and
\begin{equation}
{\cal I}_{q_1 q_2}(k_q)
=
\sum_{n=-\infty}^\infty
\delta_{q_2-q_1,q+nL}
\;.
\end{equation}
$S_{\Delta n}(k)$ takes the form~\cite{K16}
\begin{eqnarray}
S_{\Delta n}(k)
&=&
\min
\left(
    \frac{k}{2k_{\rm F}} , \xi, \frac{k_L-k}{2k_{\rm F}}
\right)
\nonumber
\\
\xi
&=&
\min
\left(
    \frac{1-f}{f}, 1
\right)
\;,
\end{eqnarray}
where $k\in[0,k_L]$.
This leads to the visible kinks in Fig.~\ref{Fig:Sk:vac}.
Note that
\begin{eqnarray}
S_{n}(k)
=
S_{\Delta n}(k)
+
N
\sum_{q=-\infty}^\infty
\delta_{k,q k_L}
\;.
\end{eqnarray}

\begin{figure}
\includegraphics[width=\columnwidth,angle=0]{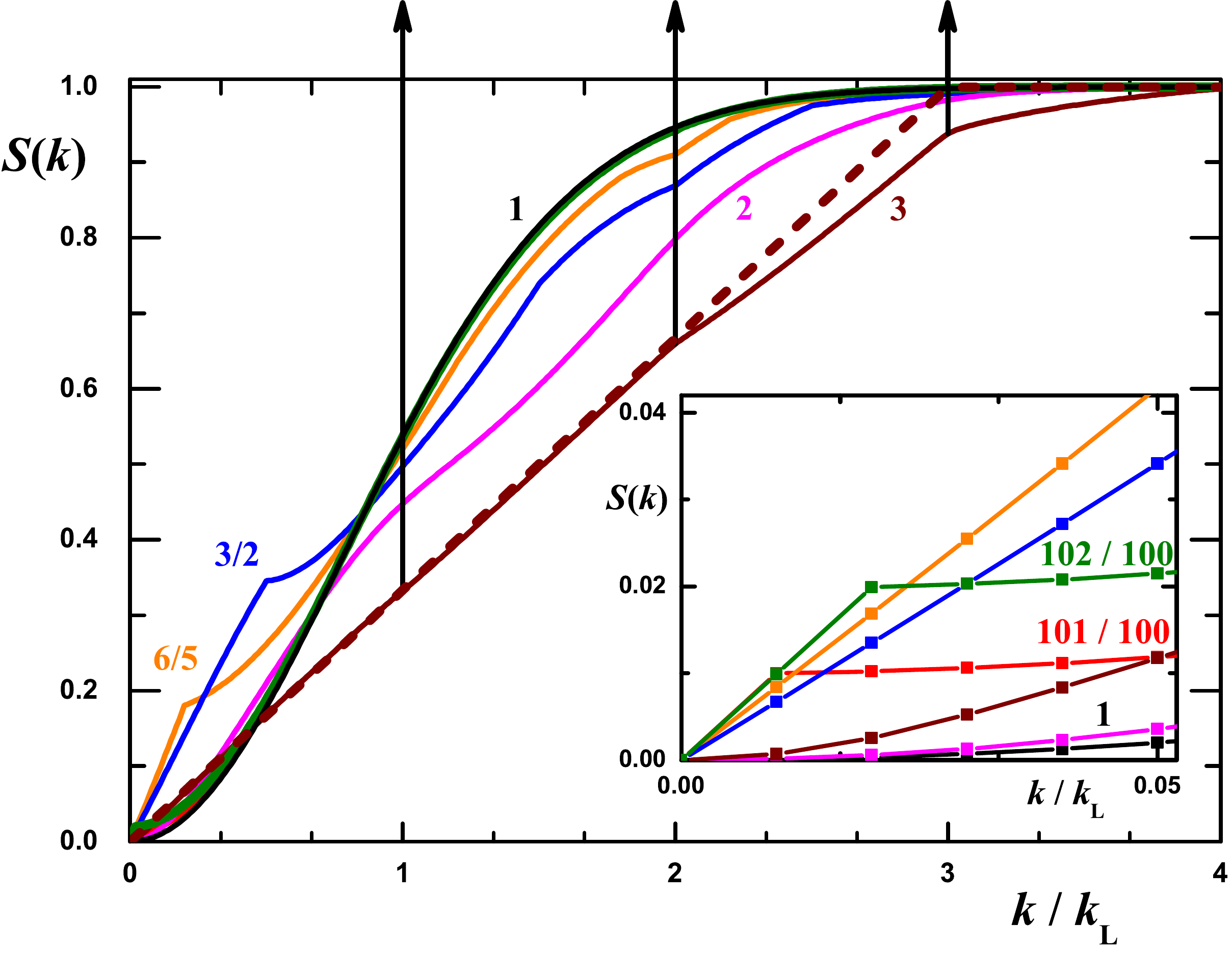}
\caption{(color online)
Static structure factor $S_{\rho}(k)$ of the Tonks-Girardeau gas in a deep lattice ($V_0 = 10\,E_{\rm rec}$) with $L=100$ lattice sites.
The fractional filling: 1; 101/100; 102/100; 6/5; 3/2; 2; 3.
Dashed line, uniform density limit, Eq.~(\ref{Eq:Sk:TG:hom}).
The peaks at commensurate momentum $k = q k_L, q = 1,2,\cdots$ are macroscopically large and are denotes with arrows and the trivial peak $S_{\rho}(0)=N$ is not shown.
Inset: zoom for small momentum part.
}
\label{Fig:Sk:int}
\end{figure}

\subsection{Discussion of numerical results}

Figure~\ref{Fig:Sk:vac} reports the static structure factor for filling smaller or equal to one.
At unit filling the system is insulating and the spectrum is gapped.
According to the Feynman relation~(\ref{Eq:Feynman}), the low-momentum behavior is quadratic with coefficient of proportional to the inverse of the gap,
$S_{\Delta\rho}(k) = \hbar^2k^2/(2m\Delta)$.
For a single vacancy it is possible to create a low-lying phonon-like excitation at the lowest allowed momentum $k_{\rm min} = 2\pi/(La_0)$, see inset in Fig.~\ref{Fig:Sk:vac}, and, accordingly, the value of $S_{\Delta\rho}(k_{\rm min})$ is largely increased.
For two vacancies it is possible to create low-lying excitations at the first two allowed momenta, both $S_{\Delta\rho}(k_{\rm min})$
and $S_{\Delta\rho}(2 k_{\rm min})$.
Eventually, for many vacancies a phononic branch $E(k)=\hbar c|k|$ is formed, leading to a linear static structure factor at low momenta $k$,
$S_{\Delta\rho}(k)=\hbar|k|/(2mc)$.
For the Tonks-Girardeau gas the speed of sound is determined by the Fermi velocity, $c=v_{\rm F}$, which is entirely defined by the density,
$v_{\rm F} = \hbar k_{\rm F} / m = \hbar \pi N / (mLa_0)$.
The slope at low $k$, in units of $k_L$, is then given by the inverse of the filling factor $f=N/L$:
\begin{eqnarray}
S_{\Delta\rho}(k)
=
\frac{|k|}{2k_{\rm F}}
=
\frac{1}{f}\frac{|k|}{k_L}
\;.
\end{eqnarray}

As anticipated at the beginning of this section, there are special values of the momenta, $k = (1-f)k_{\rm L}$,
where $S_{\Delta\rho}(k)$ shows kinks.
It is interesting to understand the underlying physical processes behind them.
In the linear regime, the upper and lower branches of excitations correspond, respectively, to the ``particle'' and ``hole''
excitatons, where the displaced particle moves outside of the Fermi sphere (particle excitation) or creates a hole in the Fermi surface (hole excitation).
Both processes are possible in a single particle excitation for momenta $0<k<2k_{\rm F}$.
Instead, for higher momentum a particle is always displaced outside of the Fermi surface.
This abrupt change in the structure of the excitations results in the kinks in the static structure factor at
$k=2k_{\rm F}=f k_L$ and $k = (1-f)k_L$.

The static structure factor for filling fraction $f\ge 1$ is shown in Fig.~\ref{Fig:Sk:int}.
For a reduced number of interstitials, only the lowest momenta get strongly affected.
Similarly to the $f<1$ case, the quadratic dependence at small momenta gets gradually replaced by a linear behavior.
When the number of interstitials grows, the linear part extends further.
As the filling fraction is increased, the static structure factor becomes more similar to that of a uniform system, given by
Eq.~(\ref{Eq:Sk:TG:hom}) with the Fermi momentum $k_{\rm F}=f k_L/2$ (see the case $f=3$ in Fig.~\ref{Fig:Sk:int}).
This is because highly-energetic states are less affected by the optical lattice.
Still, for an integer filling, there is always a gap in the excitation spectrum in the Tonks-Girardeau regime, although the value of the gap diminishes as $f$ is increased.

\section{Spatial correlations\label{Sec:g2}}

In our model the external lattice induces spatial density ordering with the period equal to the lattice spacing $a_0$.
An important issue is to determine whether the system is capable of forming a {\em spontaneous} ordering with a period different from that imposed by the external field.

\begin{figure}
\includegraphics[width=\columnwidth,angle=0]{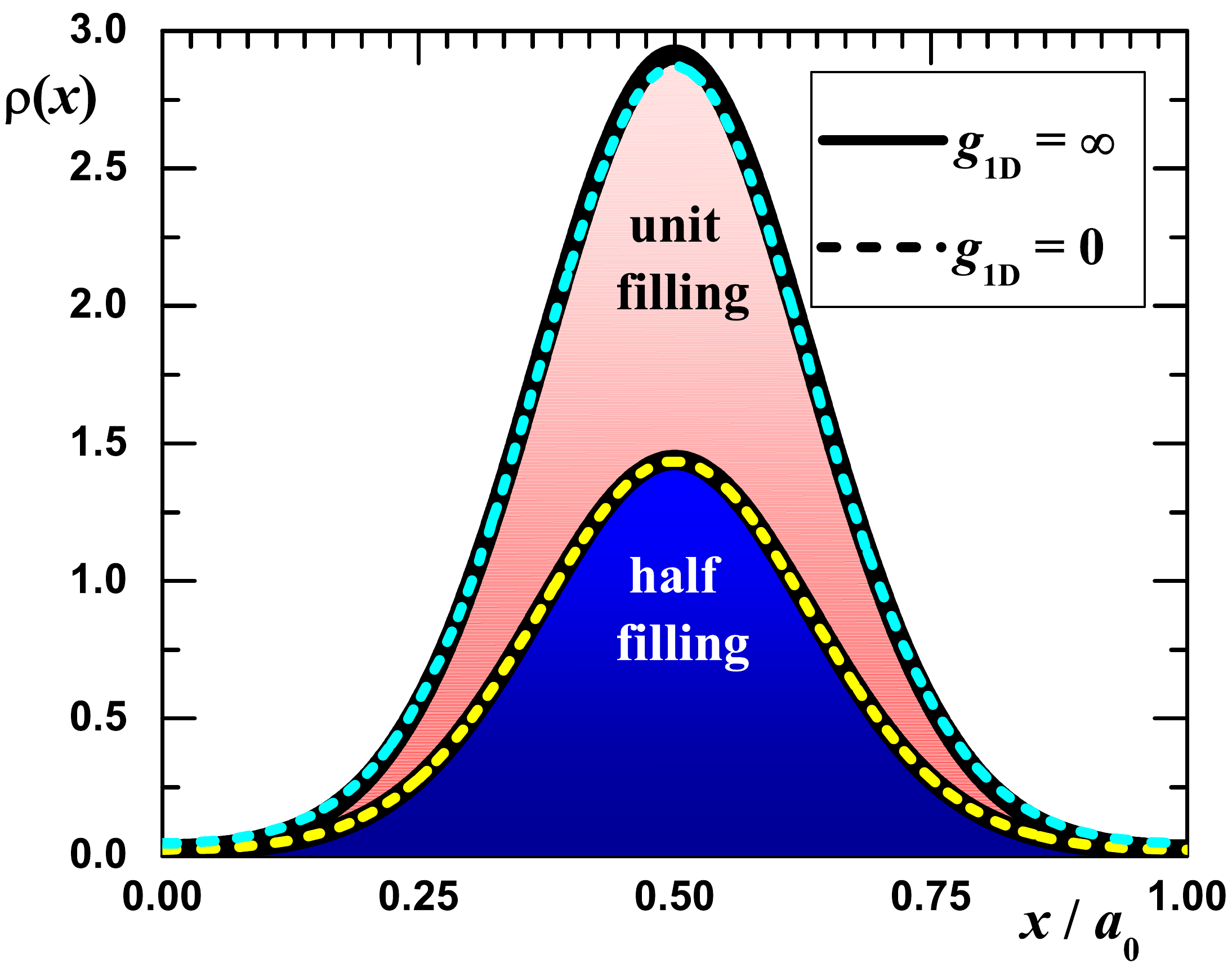}
\caption{(color online)
Density profile in a deep lattice ($V_0 = 10\,E_{\rm rec}$) for unit filling (above) and half filling (below) is shown for a single lattice period.
Density for other periods is obtained by simply repeating the shown density.
Two extreme cases of interactions are shown:
Tonks-Girardeau gas, $g_{\rm 1D} = \infty$, solid lines;
ideal Bose gas, $g_{\rm 1D} = 0$, dashed lines.
Integral of $\rho(x)$ over one period gives 1 for unit filling and 1/2 for half filling.
}
\label{Fig:density}
\end{figure}

Figure~\ref{Fig:density} shows typical examples of density profiles (only a single period is shown).
As can be observed, for deep optical lattices the strength of the interaction has a minor effect on the density (compare the cases of infinite and zero interactions).
The spreading of particles close to lattice sites is mostly controlled by the height of the optical lattice and it mimics the spreading of particles in quantum crystals.
A typical way to control the localization strength in real crystals is to change the density (or pressure), which provides a rather limited and complicated way of controlling the system parameters.
An important advantage of incommensurate gases in optical lattices as a model for defects in quantum crystals is the possibility to controllably change the particle localization and to study its effect on the superfluid response by changing the intensity of the laser beams.

In order to address the question of whether for some parameters, the system spontaneously self-organizes into a solid with a period different from that of the underlying lattice, we study a particularly well-suited situation of half filling.
This case is appropriate for a potential formation of a crystal with double spacing.
If such a crystal exists, the state with particles occupying odd sites will be degenerate with a state in which particles occupy even sites.
In that case, the equilibrium density corresponds to the average over both double-period states, with a resulting density profile where only single-period oscillations are visible (see also Fig.~\ref{Fig:density}).
Thus, the effect of statistical averaging prevents observation of the period doubling in the total density.
Instead, the double period should be still visible in the pair distribution function, which is the same for the states occupying even and odd positions.

\begin{figure}
\includegraphics[width=\columnwidth,angle=0]{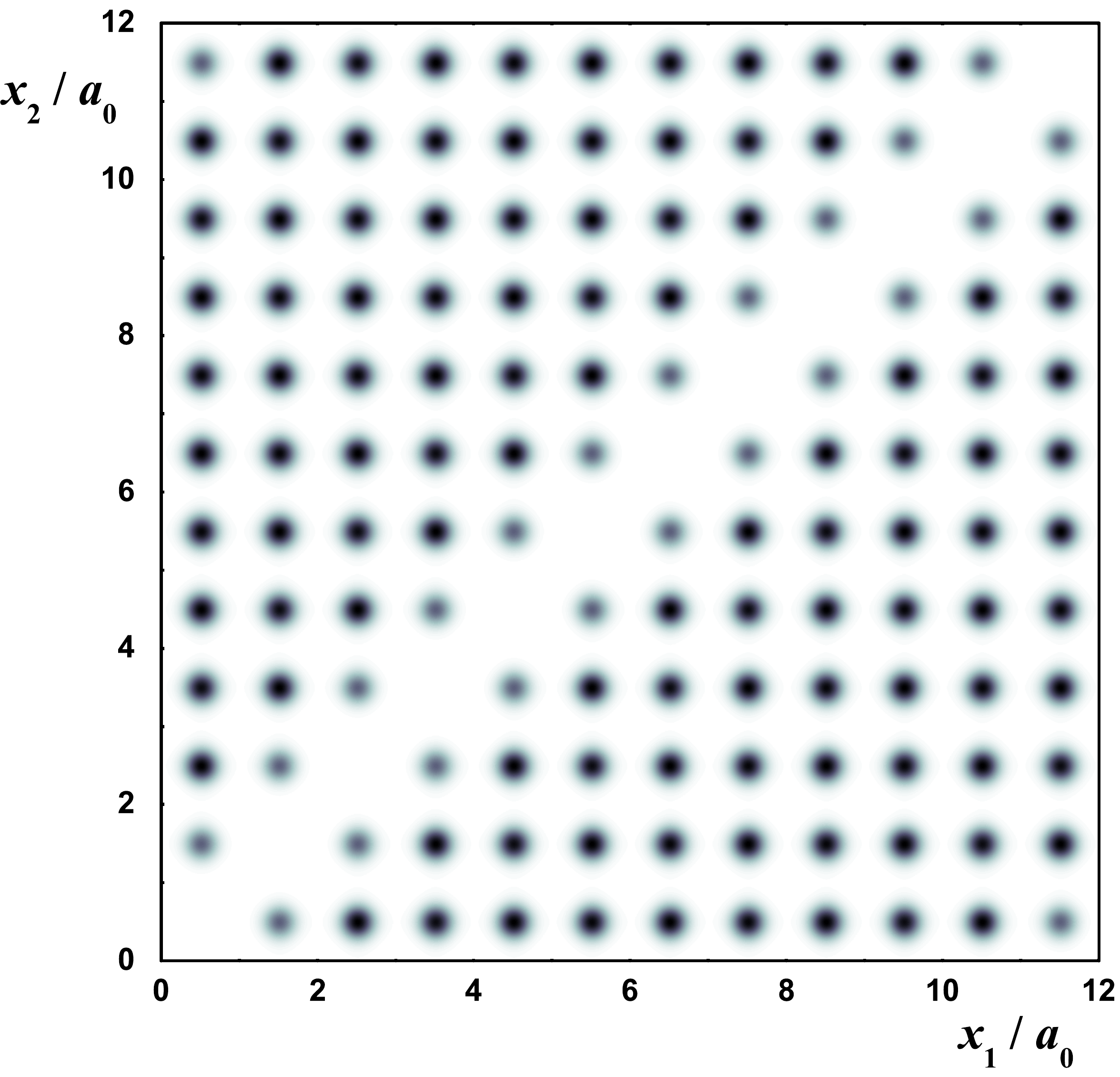}
\caption{(color online)
Pair density function in a deep lattice ($V_0 = 10\,E_{\rm rec}$) for the Tonks-Girardeau gas for $N=L=12$.
}
\label{Fig:g2matrix}
\end{figure}

Figure~\ref{Fig:g2matrix} shows an example of a pair distribution function $g_2(x_1, x_2)$ which is proportional to the probability of simultaneously finding two particles at positions $x_1$ and $x_2$.
In one dimension, $g_2(x_1, x_2)$ depends only on two scalar arguments and can be conveniently visualized with a contour plot.
In the Tonks-Girardeau regime, strong repulsion effectively prohibits double site occupation, as seen from the void diagonal in Fig.~\ref{Fig:g2matrix}.
A regular structure that follows the lattice period can be clearly seen.

\begin{figure}
\includegraphics[width=\columnwidth,angle=0]{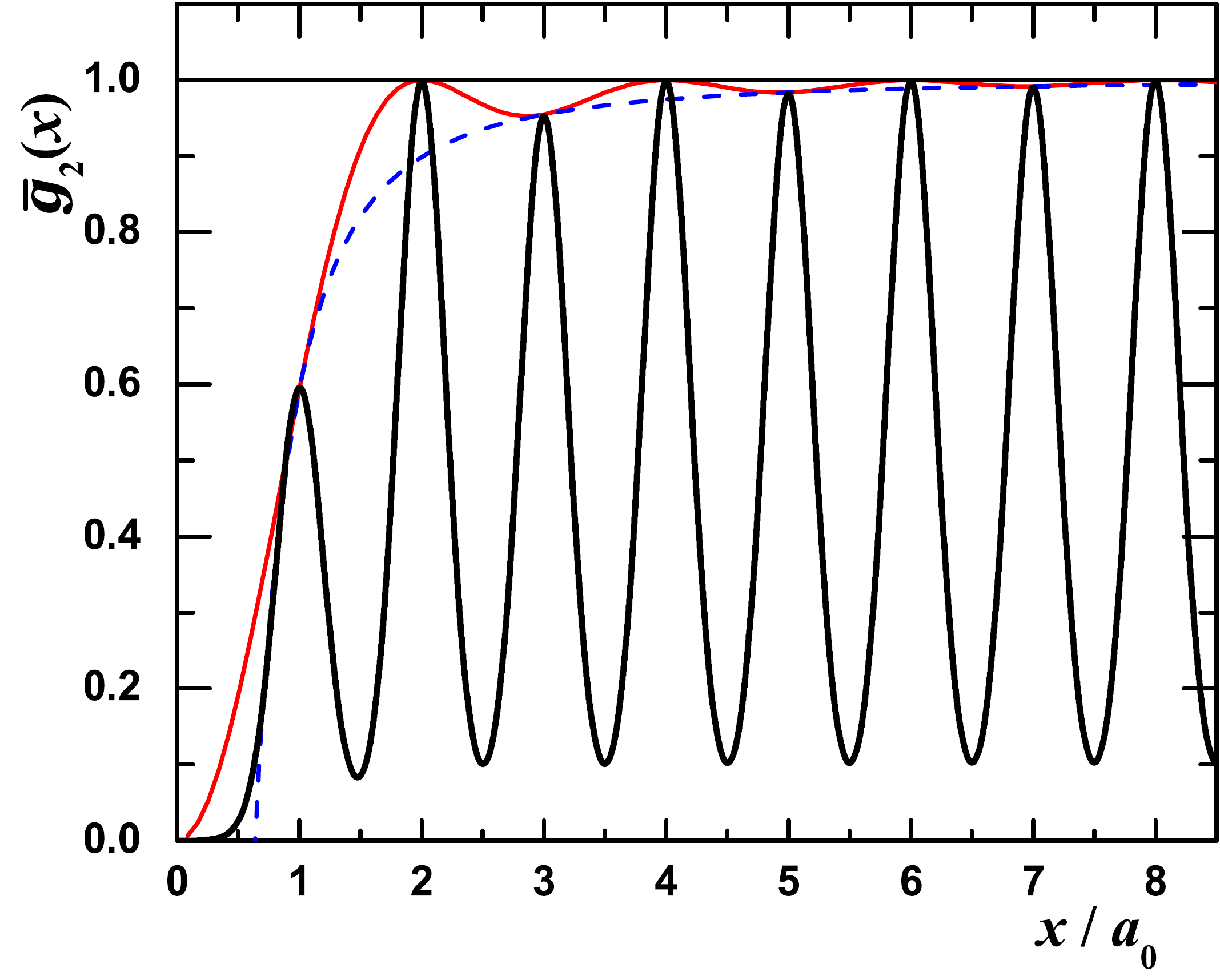}
\caption{(color online)
Thick solid black line, averaged pair distribution function as defined by Eq.~(\ref{Eq:g2average}) for the same parameters as in Fig.~\ref{Fig:g2matrix}.
Thin solid red line, ideal Fermi gas result~(\ref{Eq:g2free}) giving the height of the peaks.
Thin dashed blue line, inverse-square decay~(\ref{Eq:g2decay}).
}
\label{Fig:g2averaged}
\end{figure}

By integrating out the position of the center of mass $R = (x_1 + x_2)/2$ one effectively reduces the function to the relative coordinate $x = x_1 - x_2$,
\begin{eqnarray}
\overline{g}_2(x)
=
\frac{1}{L a_0}
\int_0^{L a_0}
g_2
\left(
    R+\frac{x}{2},R-\frac{x}{2}
\right)
\,dR
\;.
\label{Eq:g2average}
\end{eqnarray}
This averaged pair distribution function is closely related to Fourier transform of the static structure factor,
\begin{equation}
\overline{g}_2(x)
=
\frac{N}{L^2}
\sum_k
\left[
    S_\rho(k)-1
\right]
e^{ikx}
\;.
\end{equation}
Figure~\ref{Fig:g2averaged} shows $\overline{g}_2(x)$ of the Tonks-Girardeau gas at half filling.
It can be appreciated that the even peaks are higher than the odd peaks, which reflects the tendency of period doubling.
Still this tendency does not result in a {\rm long-range} order as the corresponding correlations decay as a power law.
For the comparison we show the pair-distribution function of the Tonks-Girardeau gas in the homogeneous space
\begin{eqnarray}
{\overline g}^{\rm hom}_2(x)a_0^2
=
f^2
\left[
    1 - \frac{\sin^2(\pi f x/a_0)}{(\pi f x/a_0)^2}
\right]
\;.
\label{Eq:g2free}
\end{eqnarray}
For the discrete Bose-Hubbard model it can be demonstrated~\cite{K16} that Eq.~(\ref{Eq:g2free}) describes the pair distribution function of the Tonks-Girardeau gas at the discrete positions $x=x_\ell=\ell a_0$.
As can be seen from Fig.~\ref{Fig:g2averaged}, in a continuous description the ideal Fermi gas behavior~(\ref{Eq:g2free}) is valid for the discrete points $x = \ell a_0, \ell = 0;\pm 1; \pm 2, \cdots$, that it exactly at the positions of the maxima.
This immediately provides the information on how the height of the odd peaks, $\ell = 1, 3, \cdots$, changes with the distance
\begin{eqnarray}
{\overline g}^{\rm hom}_2(\ell a_0)a_0^2
=
f^2
\left[
    1 - \frac{1}{(\pi f \ell)^2}
\right]
\;.
\label{Eq:g2decay}
\end{eqnarray}
The inverse-square decay (\ref{Eq:g2free}) is typical for the density fluctuations.
The long-range order is absent for the spontaneously formed double period as it is lost in a power-law decay,
while the long-range order is present for the single period, as imposed by the external periodic potential.
In momentum space this is reflected by the peak at $k=k_L/2$ in Fig.~\ref{Fig:Sk:vac}.
The height of this spontaneously formed peak is constant and does not change if the system size is increased.
Instead, the height of the peaks at the commensurate momenta $k=q k_L$ is given by the last term in Eq.~(\ref{Srho_peak}) and is linearly proportional to $N$.
This qualitative difference in the scaling of the height of the peak with the number of particles (macroscopic for commensurate and microscopic for incommensurate momenta) might be interpreted as a {\em tendency} to form a spontaneous diagonal long-range order rather than its {\em real} formation.

\section{Interaction between vacancies \label{Sec:g2vac}}

A conceptually important question is to determine an effective interaction potential between two vacancies.
Although it is very hard to find a complete answer to it, even a partial knowledge of the sign of the interaction only, permits us to obtain insights on the behavior of the vacancies.
Attractive vacancies will have a tendency to bunch together and might eventually make macroscopic holes in the system, while repulsive ones will be miscible with the host gas and no phase separation particles-vacancies is possible.

A direct measure of the attractive or repulsive character of the interaction is provided by the pair distribution function $\overline{g}_2(x)$ of Eq.~(\ref{Srho}), which is proportional to the probability of finding two particles separated by distance $x$.
A repulsive two-body interaction typically depletes $\overline{g}_2(x)$ at short distances, with a finite or zero value at $x=0$ depending on its strength around $x=0$.
However, the opposite situation is found when the interaction is attractive and the value of $\overline{g}(x)$ is enhanced at $x=0$.

Once thermalized, Monte Carlo simulations yield samples of ground state configurations which can be used to estimate the pair distribution functions.
In the present case where the number of particles $N$ is different from the number of sites $L$, Monte Carlo sampling can be used to determine both the pair distribution function of particles $\overline{g}(x)$ and the pair distribution function of vacancies $\overline{g}_{\rm vac}(x)$.
While the procedure for getting the former is standard, finding a good estimation of the latter can be tricky.
This is because one has to infer the vacancy position for each particle configuration sample, while in each of these particles can fluctuate considerably around site positions $x_\ell$.

In order to determine the pair distribution function of vacancies we follow the following procedure.
For a given particle configuration and a list of sites positions, we look for the particle coordinate $x_i$ and site position $x_\ell$ that minimize the distance $|x_i-x_\ell|$.
Once found, $x_i$ and $x_\ell$ are removed from their respective lists, and the procedure is repeated.
The algorithm ends when all particle coordinates have been removed, considering there are more sites than particles.
We finally identify the vacancy positions with the remaining site coordinates.
This procedure is similar to the {\em greedy} method described by Prokof'ev and Svistunov in Ref.~\cite{Prokofev05}, and the final distance obtained when removing the last particle coordinate provides a measure of the delocalization of particles in the given configuration.
Once a large set of particle configurations is processed, one has a list of vacancy positions that can be used to estimate the corresponding pair distribution function $\overline{g}_{\rm vac}(x)$.
We have checked that the outlined procedure and the method of Clark and Ceperley, based on a combinatorial minimization of the distances between particles and all possible sites~\cite{Ceperley08}, yield the same results in small systems with $N<10$ particles with one or two more sites while our approach greatly reduces the computational costs.

\begin{figure}
\includegraphics[width=\columnwidth,angle=0]{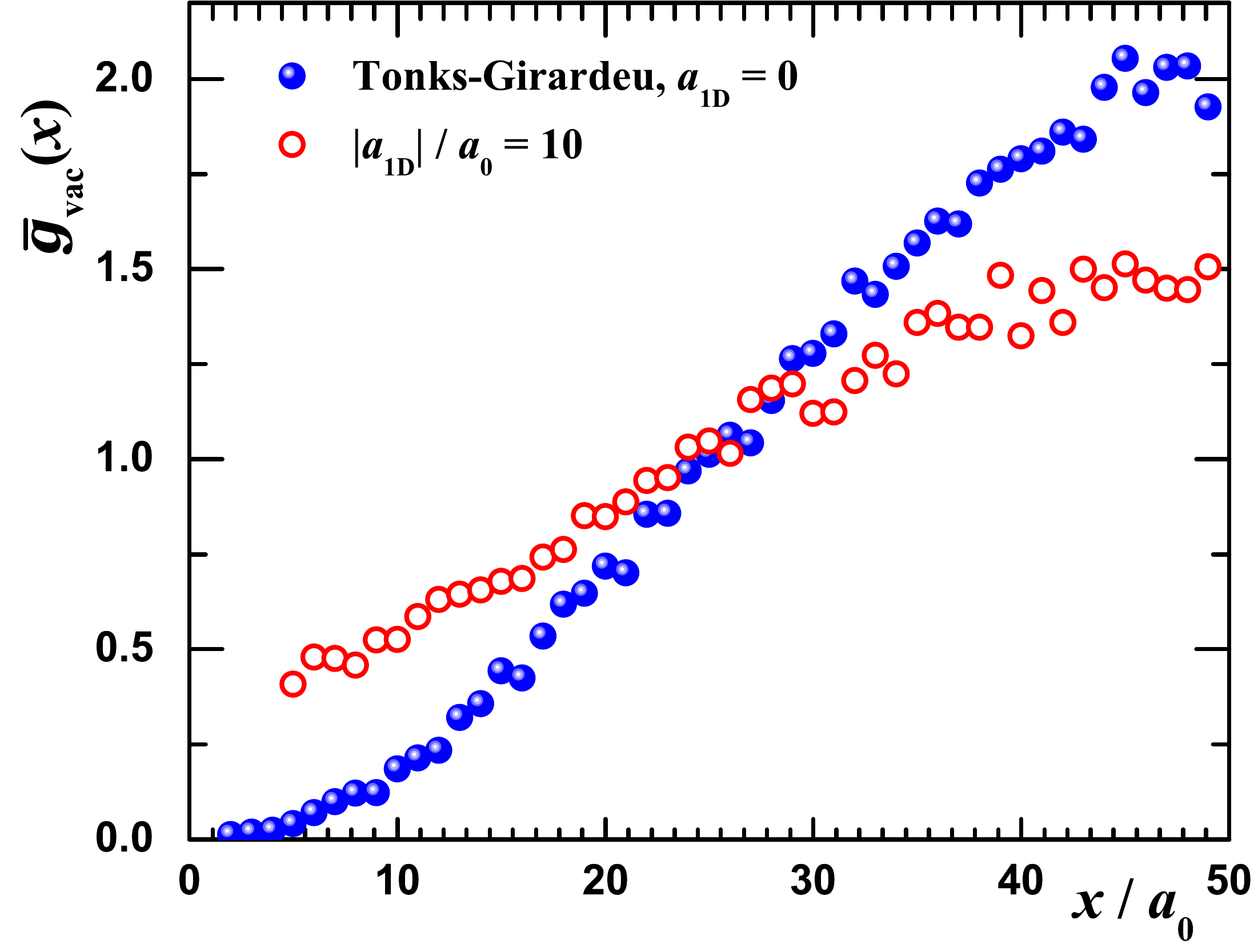}
\caption{(color online)
Pair distribution function of vacancies $\overline{g}_{\rm vac}(x)$ for the Tonks-Girardeau gas (solid line) and for the softer interaction with $\left|a_{\rm 1D}\right|/a_0=10$ (dashed line) in a lattice with $V_0=10\,E_{\rm rec}$ corresponding to $L=100$ sites and $N=98$ particles.}
\label{Fig_gv_a0_dmc_Erec10}
\end{figure}
\begin{figure}
\includegraphics[width=\columnwidth,angle=0]{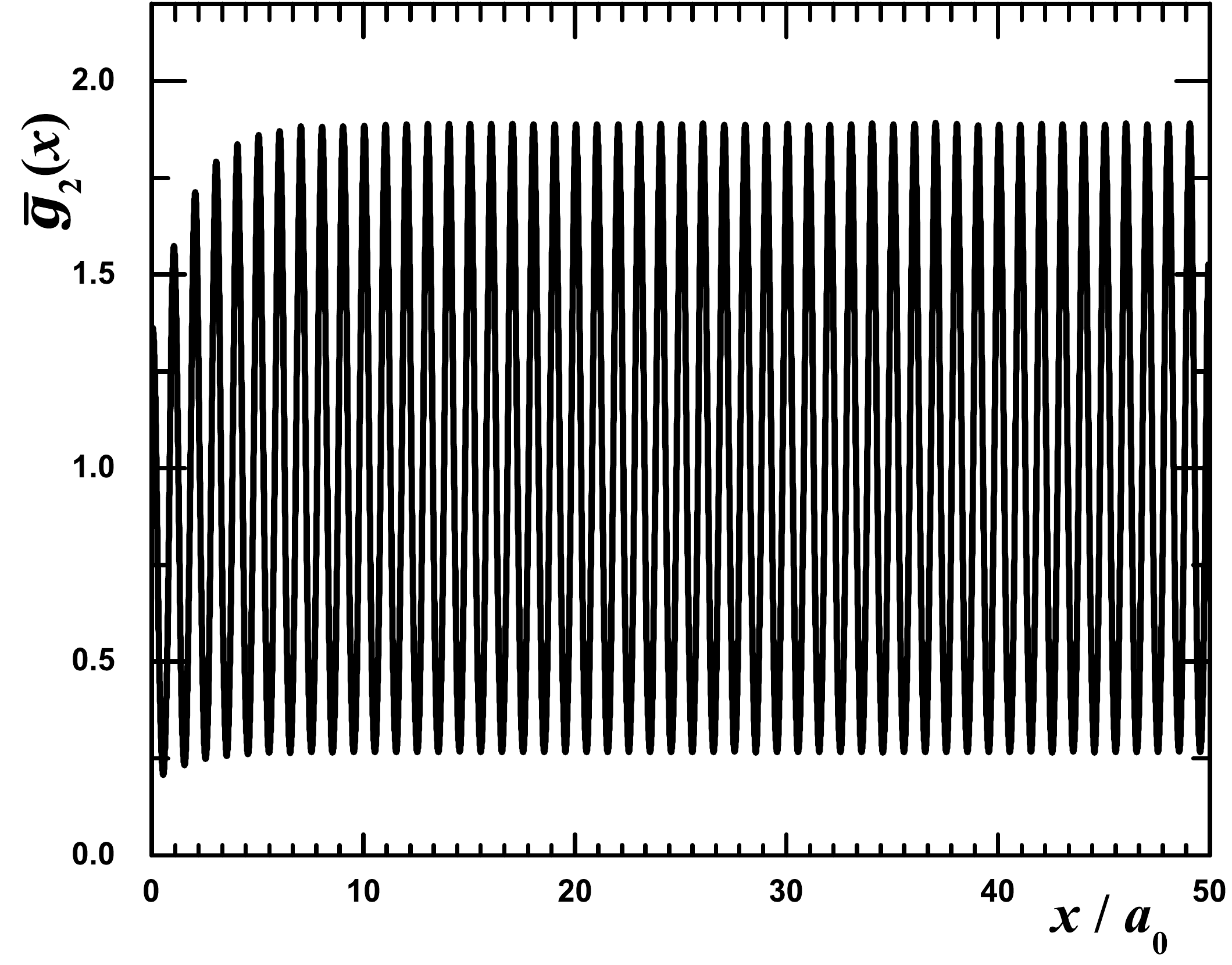}
\caption{(color online)
Pair distribution function of particles $\overline{g}(x)$ for the Tonks-Girardeau gas in a lattice with $V_0=10\,E_{\rm rec}$, $L=100$ sites and $N=98$ particles.}
\label{Fig_gx_a0_dmc_Erec10}
\end{figure}

Figure~\ref{Fig_gv_a0_dmc_Erec10} shows characteristic examples of the pair distribution function $\overline{g}_{\rm vac}(x)$ for the strong repulsion (Tonks-Girardeau gas) and for a weak interaction ($|a_{\rm 1D}| / a_0 = 10$).
Both cases correspond to two vacancies and 98 particles on a 100 sites in a deep optical lattice with $V_0=10 E_{\rm rec}$.
As it can be seen, in both cases $\overline{g}_{\rm vac}(x)$ is suppressed for short distances reaching the maximal value for the largest possible separation.
From that one concludes that, for the particle interaction employed, the effective interaction between vacancies is
repulsive, and that vacancies tend to separate as much as they can from each other.
This effect is stronger when the interaction strength between particles increases, as evidenced by the fact that the curve corresponding to $a_{\rm 1D}=0$ starts from a lower value compatible with zero at the origin, and gathers more strength than the $|a_{\rm 1D}|/a_0=10$ system at $L/2$.
We thus conclude that the effective interaction between vacancies is repulsive when the interaction between particles is also repulsive, and that the strength of the interaction between vacancies also increases as the interaction between the particles is increased.
For the sake of completeness, we show in Fig.~\ref{Fig_gx_a0_dmc_Erec10} the pair distribution function of particles $\overline{g}_2(x)$, for the same lattice and number of particles, corresponding to the Tonks-Girardeau gas.
As can be seen, the marked shell structure of this function indicates that already at this value of $V_0$ particles tend to localize around site positions, supporting the assumption that vacancies have a similar behavior.

\section{Conclusions\label{Sec:conclusions}}

To conclude, we have studied the onset of superfluidity in a one-dimensional Bose gas in optical lattices applying the language of vacancies and interstitials, commonly used in the description of supersolids.
At difference with the paradigmatic solid $^4$He case, the ground state of ultracold atoms in optical lattices can have {\em by construction} defects moving through the system as quasi-particles.
This provides a unique, highly controllable system to study how the presence of defects affect their energetic, superfluid and structural properties.
For infinite repulsion (Tonks-Girardeau limit), the Bose-Fermi mapping permits to obtain the solution of the many-particle problem in terms of the single-particle states, while for a contact $\delta$-pseudopotential of finite strength we perform quantum Monte Carlo simulations.
For deep optical lattices we employ the Bose-Hubbard model, which can be solved numerically by exact diagonalization for arbitrary interaction strength and allows simple analytical solutions in the Tonks-Girardeau limit.

When the number of defects is low, the net effect is found to be similar to that of doping in semiconductors, when each single defect in an initially insulating system contributes to the mobility (superfluidity in our case).
We verify that for a single defect the contribution to the superfluid fraction is renormalized by the ratio of the effective to the bare masses,
$N_{\rm s} = m/m_{\rm eff}$, with $m_{\rm eff}$ taking the same value as extracted from the $k \to 0$ quadratic excitation spectrum.
For an interstitial, the effective mass is $m_{\rm eff}/m>1$ and the contribution of each defect to the mobility is reduced. Instead, for a vacancy the effective mass is $m_{\rm vac}/m<1$ and each vacancy contributes with a weight larger then one.
This is a fully quantum effect stemming from enhanced quantum correlation in one-dimensional systems (as compared to three and two dimensions).

The presence of defects produces influences significantly on the structure of the excitation spectrum.
In a defect-free system there is a zero-temperature phase transition between a superfluid and a Mott insulator, with linear and gapped excitation spectra, respectively.
The presence of defects turns the gapped spectrum into a gapless one.
For a microscopic fraction of defects the low-lying excitations have a {\em quadratic} dispersion relation.
We find that this behavior is completely missed when using the Feynman approximation which, on the contrary, remains qualitatively correct in a linear (superfluid) and gapped (insulating) phases.
A macroscopic fraction of defects instead transforms a gapped spectrum into a gapless {\em linear} one.

We speculate that the interaction between vacancies is repulsive as evidenced by suppression of the vacancy-vacancy pair distribution function at short distances.
This makes phase separation into vacancy-hole regions improbable.
The interaction strength between vacancies is found to be largest for the strongest particle-particle repulsion (the Tonks-Girardeau limit).

Defects strongly modify the structural properties and particle-particle correlations, as can be seen from the changes in the static structure factor $S(k)$, which possesses a very involved shape.
Starting from the discrete Bose-Hubbard model, we show how an increase in the fraction of vacancies alters the shape of $S(k)$ in a continuous model.
In a superfluid system the low-momentum part of $S(k)$ is linear with $k$, while in an insulating system it is quadratic.
We show how the injection of vacancies introduces a linear part which can span only a limited number of $k$ values due to its finite concentration.
Finally, we note that the value of the gap $\Delta$ can be extracted from the convexity of the static structure factor which might serve as an alternative to the modulation spectroscopy.

\section{Acknowledgements}

The research leading to these results received funding from the MICINN (Spain) Grant No. FIS2014-56257-C2-1-P.
The Barcelona Supercomputing Center (The Spanish National Supercomputing Center - Centro Nacional de Supercomputaci\'on) is acknowledged for the provided computational facilities.
The authors gratefully acknowledge the Gauss Centre for Supercomputing e.V. (www.gauss-centre.eu) for funding this project by providing computing time on the GCS Supercomputer SuperMUC at Leibniz Supercomputing Centre (LRZ, www.lrz.de).
M.L. acknowledges support from Adv. ERCgrant OSYRIS, EU grant QUIC (H2020-FETPROACT-2014 No. 641122), EU STREP EQuaM, MINECO (Severo Ochoa grant SEV-2015-0522, FOQUS FIS2013-46768-P and FISICATEAMO FIS2016-79508-P), Generalitat de Catalunya (SGR 874), Fundaci\'o Privada Cellex, and CERCA Program/Generalitat de Catalunya.

\section*{Bibliography}


%

\end{document}